\documentclass[aps,twocolumn,prc,showpacs]{revtex4}
\usepackage{mathrsfs}
\usepackage{longtable,lscape}
\usepackage{txfonts}
\usepackage{amssymb}
\usepackage{indentfirst}
\usepackage{graphicx,,booktabs}
\usepackage{color}
\usepackage{amssymb}
\usepackage{epsfig}

\newcommand{\vsig}{\mbox{\boldmath$\sigma$\unboldmath}}
\newcommand{\veps}{\mbox{\boldmath$\epsilon$\unboldmath}}

\newcommand{\be}{\begin{equation}}
\newcommand{\ee}{\end{equation}}
\newcommand{\bea}{\begin{eqnarray}}
\newcommand{\eea}{\end{eqnarray}}
\newcommand{\bean}{\begin{eqnarray*}}
\newcommand{\eean}{\end{eqnarray*}}

\newcommand{\gapproxeq}{\lower
.7ex\hbox{$\;\stackrel{\textstyle >}{\sim}\;$}}
\newcommand{\lapproxeq}{\lower
.7ex\hbox{$\;\stackrel{\textstyle <}{\sim}\;$}}

\begin{document}

\title{\textbf{ Spectrum and electromagnetic transitions of bottomonium}}
\author{
Wei-Jun Deng, Hui Liu, Long-Cheng Gui~\footnote {E-mail: guilongcheng@ihep.ac.cn} and Xian-Hui Zhong
\footnote {E-mail: zhongxh@hunnu.edu.cn} }  \affiliation{ 1) Department
of Physics, Hunan Normal University,  Changsha 410081, China }

\affiliation{ 2) Synergetic Innovation
Center for Quantum Effects and Applications (SICQEA), Changsha 410081,China}

\affiliation{ 3) Key Laboratory of
Low-Dimensional Quantum Structures and Quantum Control of Ministry
of Education, Changsha 410081, China}


\begin{abstract}
Stimulated by the exciting progress in the observation of
new bottomonium states, we study the bottomonium spectrum.
To calculate the mass spectrum, we adopt a nonrelativistic screened
potential model. The radial Schr\"{o}dinger equation is solved with
the three-point difference central method, where the spin-dependent
potentials are dealt with nonperturbatively. With this treatment,
the corrections of the spin-dependent potentials to the
wave functions can be included successfully. Furthermore, we
calculate the electromagnetic transitions of the $nS$ ($n\leq 4$),
$nP$ ($n\leq 3$), and $nD$ ($n\leq 2$) bottomonium states with
a nonrelativistic electromagnetic transition operator widely applied
to meson photoproduction reactions.
Our predicted masses, hyperfine and fine splittings, electromagnetic
transition widths and branching ratios of the bottomonium states are in good agreement with
the available experimental data. In particular, the EM transitions
of $\Upsilon(3S)\to \chi_{b1,2}(1P)\gamma$, which were not well
understood in previous studies, can be reasonably explained by considering
the corrections of the spin-dependent interactions to the wave functions. We also discuss the observations
of the missing bottomonium states by using radiative
transitions. Some important radiative decay chains involving the
missing bottomonium states are suggested to be observed.
We hope our study can provide some useful references to observe
and measure the properties of bottomonium mesons in forthcoming experiments.
\end{abstract}


\maketitle

\section{Introduction}

Heavy quarkonium is considered to be an excellent laboratory to study
quantum chromodynamics (QCD) at low energies~\cite{Garmash:2015xea,Godfrey:1985xj,Eichten:1978tg}.
Due to a large mass of the heavy bottom quark, the
bottomonium system is essentially nonrelativistic, which makes it
relatively easy for us to study the perturbative and
nonperturbative QCD via the bottomonium spectroscopy with a
nonrelativistic approximation. In the past few years, great progress
has been achieved in the study of the bottomonium
spectroscopy~\cite{Eichten:2007qx,Brambilla:2004wf,Brambilla:2010cs,Bevan:2014iga}.
A fairly abundant bottomonium spectroscopy has been established in
experiments~\cite{PDG}(see Tab.~\ref{tabs}). Furthermore, many new
experiments are being and/or to be carried out at LHC and Belle.
In near future, more missing bottomonium states will be discovered
and more decay channels will be observed in experiments. Thus, it is
necessary to carry out a comprehensive study of the bottomonium
states according to the recent progress. On the one hand we can
obtain more knowledge of bottomonium states from experimental observations. On the
other hand, the predicted properties can provide some useful
references for our search for the missing bottomonium states in experiments.

In the past years, stimulated by the exciting progress in experiments,
many theoretical studies of bottomonium spectrum have been carried
out with different methods, such as the widely used potential
models~\cite{Gupta:1984jb,Kwong:1988ae,Ebert:2003,Chao:2009,Godfrey:2015dia,Wei-Zhao:2013sta,
Akbar:2015evy,Segovia:2016xqb,Barducci:2016wze},
lattice QCD~\cite{Becirevic:2014rda,Hughes:2015dba,Lewis:2011ti,Baker:2015xma},
effective Lagrangian approach~\cite{DeFazio:2008xq}, nonrelativistic
effective field theories of
QCD~\cite{Brambilla:2005zw,Brambilla:2012be,Pineda:2013lta}, various coupled-channel
quark models~\cite{Ferretti:2013vua,Ferretti:2014xqa,Lu:2016mbb}, and light front quark model~\cite{
Choi:2007se,Ke:2010vn,Ke:2010tk,Ke:2013zs}. Although some comparable
predictions from different models have been achieved, many properties of
the bottomonium states are still not well understood.
For example, the recent calculations with
the relativized quark model~\cite{Godfrey:2015dia} obtain a successful
description of the masses for the low-lying excitations, however, the
predicted mass for the higher excitation $\Upsilon(6S)$ is about 100
MeV higher than the data if $\Upsilon(11020)$ is identified as $\Upsilon(6S)$;
while the recent nonrelativistic
constituent quark model~\cite{Segovia:2016xqb} gives a good
description of the mass of $\Upsilon(6S)$, however,
the predicted masses for the ground states $\Upsilon(1S)$ and $\eta_b(1S)$
are about 50 MeV larger than the experimental values. Furthermore,
there are puzzles in the electromagnetic (EM) transitions of bottomonium states.
For example, about the M1 transitions of $\Upsilon(2S,3S)\to \eta_b(1S)\gamma$,
the predictions from the relativistic quark model~\cite{Ebert:2003}
and nonrelativistic effective field theories of
QCD~\cite{Pineda:2013lta} are about an order of magnitude smaller
than the recent predictions from the relativized quark
model~\cite{Godfrey:2015dia} and nonrelativistic constituent quark
model~\cite{Segovia:2016xqb}; while about the EM transitions
of $\Upsilon(3S)\to \chi_{b1,2}(1P)\gamma$, the predicted
partial widths in the literature~\cite{Chao:2009,Godfrey:2015dia,Segovia:2016xqb}
are inconsistent with the data. Thus, to deepen our knowledge about
the bottomonium spectrum, more theoretical studies are
needed.

In this work, first we use the nonrelativistic screened potential
model~\cite{Li:2009zu,Chao:2009,ChaoKT90,ChaoKT93} to calculate the masses and
wave functions. In this
model, the often used linear potential $br$ is replaced with the
screened potential $b(1-e^{-\mu r})/\mu$. The reason is that the
linear potential, which is expected to be dominant at large
distances, is screened or softened by the vacuum polarization effect
of the dynamical light quark
pairs~\cite{Laermann:1986pu,Born:1989iv}. Such a screening effect
might be important for us to reasonably describe the higher radial
and orbital excitations. Considering the corrections of
the spin-dependent interactions to the space wave functions cannot
be included with the perturbative treatment, we treat the
spin-dependent interactions as nonperturbations in our
calculations. With the nonperturbative treatment, we can reasonably
include the effect of spin-dependent interactions on the
wave functions, which is important for us to gain reliable predictions
of the decays.


Moreover, using the obtained wave functions, we study the EM
transitions between bottomonium states. Difference of our method
from the often used potential models is that the EM transition
operator between initial and final hadron states is used a special
nonrelativistic form
$h_{e}\simeq\sum_{j}[e_{j}\mathbf{r}_{j}\cdot\veps-\frac{e_{j}}{2m_{j}
}\vsig_{j}\cdot(\veps\times\hat{\mathbf{k}})]e^{-i\mathbf{k}\cdot
\mathbf{r}_{j}}$~\cite{Deng:2015bva}, which has been well developed and widely applied
to meson photoproduction reactions~\cite{Li:1994cy,Li:1995si,Li:1997gda,
Zhao:2001kk, Saghai:2001yd, Zhao:2002id, He:2008ty, He:2008uf,
He:2010ii, Zhong:2011ti, Zhong:2011ht, Xiao:2015gra}. In this
operator, the effect of binding potential between quarks is
considered. Furthermore, the possible higher EM multipole
contributions to a EM transition process can be included naturally.

The paper is organized as follows. In Sec.~\ref{spectrum}, we
calculate the masses and wave functions within a screened potential model.
In Sec.~\ref{EMT}, the EM transitions between the bottomonium states are calculated, and
our analysis and discussion are given.
Finally, a summary is given in Sec.~\ref{sum}.

\begin{table}[htb]
\begin{center}
\caption{Predicted masses (MeV) of bottomonium states. For comparison, the
measured masses (MeV) from the PDG~\cite{PDG}, and the theoretical
predictions with the previous screened potential model (SNR model)~\cite{Chao:2009},
relativized quark model (GI model)~\cite{Godfrey:2015dia}, and  nonrelativistic
constituent quark model (NR model)~\cite{Segovia:2016xqb} are also listed in
the same table. }\label{tabs}
\begin{tabular}{ccccccccc}
\hline\hline
 $n^{2S+1}L_J$ & name &$J^{PC}$ &PDG~\cite{PDG}  &SNR\cite{Chao:2009} &GI\cite{Godfrey:2015dia}& NR\cite{Segovia:2016xqb}& Ours\\
 \hline
$1 ^3S_{1}$     &  $\Upsilon(1S)$            &$1^{--}$    &$9460$       &$9460$       &9465 &9502 &9460 \\
$1 ^1S_{0}$     &  $\eta_{b}(1S)$            &$0^{-+}$    &$9398$       &$9389$       &9402 &9455 &9390 \\
$2 ^3S_{1}$     &  $\Upsilon(2S)$            &$1^{--}$    &$10023$      &$10016$      &10003&10015&10015   \\
$2 ^1S_{0}$     &  $\eta_{b}(2S)$            &$0^{-+}$    &$9999$       &$9987$       &9976 &9990 &9990 \\
$3 ^3S_{1}$     &  $\Upsilon(3S)$            &$1^{--}$    &$10355$      &$10351$      &10354&10349&10343   \\
$3 ^1S_{0}$     &  $\eta_{b}(3S)$            &$0^{-+}$    &             &$10330$      &10336&10330&10326   \\
$4 ^3S_{1}$     &  $\Upsilon(4S)$            &$1^{--}$    &$10579$      &$10611$      &10635&10607&10597   \\
$4 ^1S_{0}$     &  $\eta_{b}(4S)$            &$0^{-+}$    &             &$10595$      &10623&     &10584   \\
$5 ^3S_{1}$     &  $\Upsilon(5S)$            &$1^{--}$    &$10865$      &$10831$      &10878&10818&10811   \\
$5 ^1S_{0}$     &  $\eta_{b}(5S)$            &$0^{-+}$    &             &$10817$      &10869&     &10800   \\
$6 ^3S_{1}$     &  $\Upsilon(6S)$            &$1^{--}$    &$11020$      &$11023$      &11102&10995&10997  \\
$6 ^1S_{0}$     &  $\eta_{b}(6S)$            &$0^{-+}$    &             &$11011$      &11097&     &10988   \\
$1 ^3P_{2}$     &  $\chi_{b2}(1P)$           &$2^{++}$    &$9912$       &$9918$       &9897 &9886 & 9921\\
$1 ^3P_{1}$     &  $\chi_{b1}(1P)$           &$1^{++}$    &$9893$       &$9897$       &9876 &9874 & 9903\\
$1 ^3P_{0}$     &  $\chi_{b0}(1P)$           &$0^{++}$    &$9859$       &$9865$       &9847 &9855 & 9864\\
$1 ^1P_{1}$     &  $h_{b}(1P)$               &$1^{+-}$    &$9899$       &$9903$       &9882 &9879 &9909\\
$2 ^3P_{2}$     &  $\chi_{b2}(2P)$           &$2^{++}$    &$10269$      &$10269$      &10261&10246& 10264  \\
$2 ^3P_{1}$     &  $\chi_{b1}(2P)$           &$1^{++}$    &$10255$      &$10251$      &10246&10236& 10249  \\
$2 ^3P_{0}$     &  $\chi_{b0}(2P)$           &$0^{++}$    &$10233$      &$10226$      &10226&10221& 10220  \\
$2 ^1P_{1}$     &  $h_{b}(2P)$               &$1^{+-}$    &$10260$      &$10256$      &10250&10240& 10254  \\
$3 ^3P_{2}$     &  $\chi_{b2}(3P)$           &$2^{++}$    &             &$10540$      &10550&10521   &10528\\
$3 ^3P_{1}$     &  $\chi_{b1}(3P)$           &$1^{++}$    &10516        &$10524$      &10538&10513   &10515\\
$3 ^3P_{0}$     &  $\chi_{b0}(3P)$           &$0^{++}$    &             &$10502$      &10522&10500   &10490\\
$3 ^1P_{1}$     &  $h_{b}(3P)$               &$1^{+-}$    &             &$10529$      &10541&10516   &10519\\
$1 ^3D_{3}$     &  $\Upsilon_3(1D)$          &$3^{--}$    &             &$10156$      &10155&10127&10157   \\
$1 ^3D_{2}$     &  $\Upsilon_2(1D)$          &$2^{--}$    &$10164$      &$10151$      &10147&10122&10153   \\
$1 ^3D_{1}$     &  $\Upsilon_1(1D)$          &$1^{--}$    &             &$10145$      &10138&10117&10146   \\
$1 ^1D_{2}$     &  $\eta_{b2}(1D)$           &$2^{-+}$    &             &$10152$      &10148&10123& 10153  \\
$2 ^3D_{3}$     &  $\Upsilon_3(2D)$          &$3^{--}$    &             &$10442$      &10455&10422&10436\\
$2 ^3D_{2}$     &  $\Upsilon_2(2D)$          &$2^{--}$    &             &$10438$      &10449&10418&10432\\
$2 ^3D_{1}$     &  $\Upsilon_1(2D)$          &$1^{--}$    &             &$10432$      &10441&10414   &10425\\
$2 ^1D_{2}$     &  $\eta_{b2}(2D)$           &$2^{-+}$    &             &$10439$      &10450&10419   &10432\\
$1 ^1F_{3}$     &  $h_{b3}(1F)$              &$3^{+-}$    &             &             &10355&10322   & 10339  \\
$1 ^3F_{4}$     &  $\chi_{b4}(1F)$           &$4^{++}$    &             &             &10358&        &10340\\
$1 ^3F_{3}$     &  $\chi_{b3}(1F)$           &$3^{++}$    &             &             &10355&10321   &10340\\
$1 ^3F_{2}$     &  $\chi_{b2}(1F)$           &$2^{++}$    &             &             &10350&10315   &10338\\
\hline\hline
\end{tabular}
\end{center}
\end{table}

\begin{figure}[ht] \centering \epsfxsize=8.6 cm \epsfbox{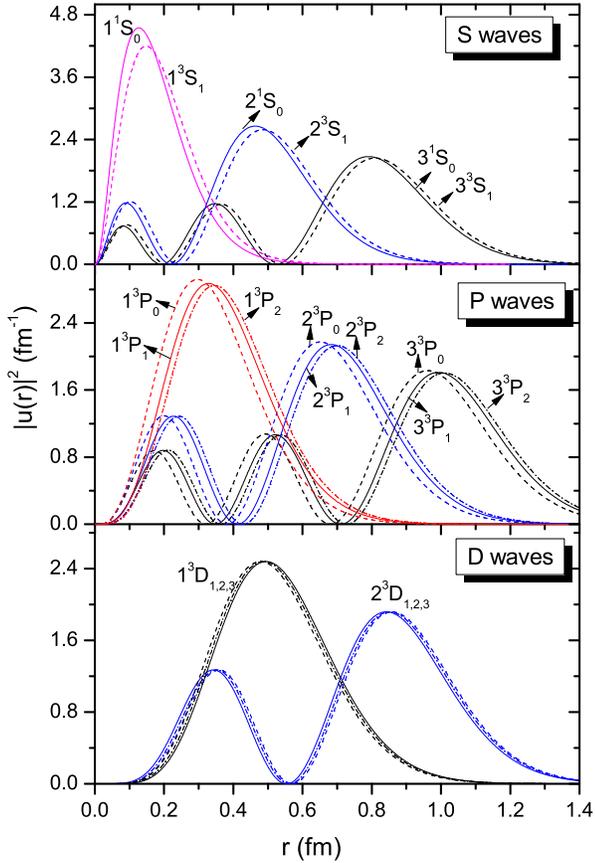}
\vspace{-1.5cm} \caption{(Color online) Predicted radial probability density
$|u(r)|^2$ for $S$-, $P$- and $D$-wave bottomonium states.
}\label{wf}
\end{figure}

\begin{table}[htb]
\begin{center}
\caption{ Hyperfine and fine splittings in units of MeV for bottomonium in our calculation.
The experimental data are taken from the PDG~\cite{PDG}. The theoretical
predictions with the previous screened potential model~\cite{Chao:2009},
relativized quark model~\cite{Godfrey:2015dia},
relativistic two-body calculation~\cite{Barducci:2016wze}, and nonrelativistic
constituent quark model~\cite{Segovia:2016xqb,Wei-Zhao:2013sta} are also listed in
the same table for comparison. }\label{tabhf}
\begin{tabular}{c|cccccccc}
\hline\hline
 $\Delta m$ & Ours &~\cite{Chao:2009} &~\cite{Godfrey:2015dia}  &~\cite{Segovia:2016xqb}  &\cite{Barducci:2016wze}&\cite{Wei-Zhao:2013sta}&PDG~\cite{PDG} \\
 \hline
$m(1 ^3S_{1})$-$m(1 ^1S_{0})$    &  70            &71    &63       &47      &76 &49  &$62.3\pm 3.2$ \\
$m(2 ^3S_{1})$-$m(2 ^1S_{0})$    &  25            &29    &27       &25      &38 &16  &$24.3^{+4.0}_{-4.5}$ \\
$m(3 ^3S_{1})$-$m(3 ^1S_{0})$    &  17            &21    &18       &19      &   &11  &  \\
$m(4 ^3S_{1})$-$m(4 ^1S_{0})$    &  13            &16    &12       &        &   &11  & \\
$m(5 ^3S_{1})$-$m(5 ^1S_{0})$    &  11            &14    &9        &        &   &15  & \\
$m(6 ^3S_{1})$-$m(6 ^1S_{0})$    &  9             &12    &5        &        &   &    & \\
$m(1 ^3P_{2})$-$m(1 ^3P_{1})$    &  18            &21    &21       &12      &22 &18  &$19.43\pm 0.57$ \\
$m(1 ^3P_{1})$-$m(1 ^3P_{0})$     & 39            &32    &29       &19      &29 &48  &$33.34\pm 0.66$ \\
$m(2 ^3P_{2})$-$m(2 ^3P_{1})$     & 15            &18    &15       &10      &18 &16  &$13.19\pm 0.77$ \\
$m(2 ^3P_{1})$-$m(2 ^3P_{0})$     & 29            &25    &20       &15      &24 &40  &$22.96\pm 0.84$ \\
$m(3 ^3P_{2})$-$m(3 ^3P_{1})$     & 13            &16    &12       &8       &  &14   &\\
$m(3 ^3P_{1})$-$m(3 ^3P_{0})$     & 25            &22    &16       &13      &  &36   & \\
\hline\hline
\end{tabular}
\end{center}
\end{table}

\section{mass spectrum}\label{spectrum}

As a minimal model of the bottomonium system we use a
nonrelativistic screened potential model~\cite{Li:2009zu,Chao:2009,ChaoKT90,ChaoKT93}.
The effective potential of spin-independent term $V(r)$ is regarded as
the sum of Lorentz vector $V_V(r)$ and Lorentz scalar $V_s(r)$
contributions~\cite{Eichten:2007qx}, i.e.,
\begin{eqnarray}\label{vr}
V(r)=V_V(r)+V_s(r).
\end{eqnarray}
For the Lorentz vector potential $V_V(r)$, we adopt the standard color
Coulomb form:
\begin{eqnarray}\label{vv}
V_V(r)=-\frac{4}{3}\frac{\alpha_s}{r}.
\end{eqnarray}
To take into account the screening effects, which might originate from the
vacuum polarization of the dynamical light quark
pairs~\cite{Laermann:1986pu,Born:1989iv}, we replace the widely used linear scalar
potential $br$ with a special form
\begin{eqnarray}\label{vs}
V_s(r)= \frac{b(1-e^{-\mu r})}{\mu},
\end{eqnarray}
as suggested in Refs.~\cite{Li:2009zu,Chao:2009,ChaoKT90,ChaoKT93}. Here $\mu$
is the screening factor which makes the long-range scalar potential
of $V_s(r)$ behave like $br$ when $r\ll 1/\mu$, and become a
constant $b/\mu$ when $r\gg 1/\mu$. The main effect of the screened
potential on the spectrum is that the masses of the higher excited
states are lowered. Such a screening effect
might be important for us to reasonably describe the higher radial
and orbital excitations.

We include three spin-dependent potentials as follows. For the
spin-spin contact hyperfine potential, we take~\cite{Barnes:2005pb}
\begin{eqnarray}\label{ss}
H_{SS}= \frac{32\pi\alpha_s}{9m_b^2}\tilde{\delta}_\sigma(r)\mathbf{S}_b\cdot \mathbf{S}_{\bar{b}},
\end{eqnarray}
where $\mathbf{S}_b$ and $\mathbf{S}_{\bar{b}}$  are spin matrices acting on the spins
of the quark and antiquark. We take $\tilde{\delta}_\sigma(r)=(\sigma/\sqrt{\pi})^3
e^{-\sigma^2r^2}$ as in Ref.~\cite{Barnes:2005pb}.
The five parameters in the above equations ($\alpha_s$, $b$, $\mu$, $m_b$, $\sigma$)
are determined by fitting the spectrum.

For the spin-orbit term and the tensor term, we take the common
forms~\cite{Eichten:2007qx}:
\begin{eqnarray}\label{sl}
H_{SL}= \frac{1}{2m_b^2r}\left(3\frac{dV_V}{dr}-\frac{dV_s}{dr}\right)\mathbf{L}\cdot \mathbf{S},
\end{eqnarray}
and
\begin{eqnarray}\label{t}
H_{T}= \frac{1}{12m_b^2}\left(\frac{1}{r}\frac{dV_V}{dr}-\frac{d^2V_V}{dr^2}\right)S_T,
\end{eqnarray}
where $\mathbf{L}$ is the relative orbital angular momentum of $b$ and $\bar{b}$ quarks,
$\mathbf{S}=\mathbf{S}_b+\mathbf{S}_{\bar{b}}$ is the total quark spin, and the
spin tensor $S_T$ is defined by~\cite{Eichten:2007qx}
\begin{eqnarray}\label{st}
S_T= 6\frac{\mathbf{S}\cdot \mathbf{r}\mathbf{S}\cdot \mathbf{r}}{r^2}-2\mathbf{S}^2.
\end{eqnarray}
In the $|^{2S+1}L_J\rangle$ basis, the matrix element for the spin-spin operator
$\mathbf{S}_b\cdot \mathbf{S}_{\bar{b}}$ is
\begin{eqnarray}\label{mss}
\langle \mathbf{S}_b\cdot \mathbf{S}_{\bar{b}}\rangle &= &\frac{1}{2}S(S+1)-\frac{3}{4}.
\end{eqnarray}
For the spin-orbit operator $\mathbf{L}\cdot \mathbf{S}$, its matrix element is
\begin{eqnarray}\label{msl}
\langle\mathbf{L}\cdot \mathbf{S}\rangle &=&\frac{1}{2}[J(J+1)-L(L+1)-S(S+1)].
\end{eqnarray}
The element of the tensor operator $S_T$ can be written in the form~\cite{DF1982}
\begin{eqnarray}\label{mtt}
\langle S_T\rangle &=&\frac{4\langle \mathbf{S}^2\mathbf{L}^2 -\frac{3}{2}\mathbf{L}
\cdot \mathbf{S}-3 (\mathbf{L}\cdot \mathbf{S})^2 \rangle}{(2L+3)(2L-1)}.
\end{eqnarray}

To obtain masses and wave functions of the bottomonium states,
we need to solve the radial Schr\"{o}dinger equation
\begin{eqnarray}\label{sdg}
\frac{d^2u(r)}{dr^2}+2\mu_R \left[E-V_{b\bar{b}}(r)-\frac{L(L+1)}{2\mu_R r^2}\right]u(r)=0,
\end{eqnarray}
with
\begin{eqnarray}\label{mtt}
V_{b\bar{b}}(r)=V(r)+H_{SS}+H_{SL}+H_{T},
\end{eqnarray}
where $\mu_R=m_b m_{\bar{b}}/(m_b+m_{\bar{b}})$ is the reduced mass
of the system, and $E$ is the binding energy of the system.
Then, the mass of a $b\bar{b}$ state is obtained by
\begin{eqnarray}\label{mtt}
M_{b\bar{b}}=2m_b+E.
\end{eqnarray}
In the literature, the spin-dependent
interactions were usually dealt with perturbatively. Although the meson mass
obtains perturbative corrections from these spin-dependent
potentials, the wave functions obtain no corrections from these
spin-dependent potentials. To reasonably include the corrections
from these spin-dependent potentials to both the mass and wave
function of a meson state, we deal with the spin-dependent
interactions nonperturbatively.

In this work, we solve the radial Schr\"{o}dinger equation by using
the three-point difference central method~\cite{Haicai} from central ($r=0$)
towards outside ($r\to \infty$) point by point.
In this method, we need to know the role of $u(r\to 0)$.
When $r\to 0$ we easily obtain $u(r\to 0)\propto
r^{L+1}$ if we neglect the contributions of the spin-orbit and tensor terms.
However, including the spin-orbit and tensor potential
contributions, we have a term $\propto 1/r^3$ in the potential. In the limit
$r\to 0$, the potential $V_{b\bar{b}}(r)\propto 1/r^3$. In this
case, we do not know the role of $u(r\to 0)$, thus, we cannot solve
the radial Schr\"{o}dinger equation with the three-point
differential central method. To overcome this problem, we assume
that in a small range $r\in (0,r_c)$, the $V_{b\bar{b}}(r)\propto
1/r_c^3$, which is a finite constant. Then, the role of $u(r\to 0)$
is still $\propto r^{L+1}$. The price of our method is that a cutoff distance
$r_c$ should be introduced in the calculation, which is determined
by fitting the spectrum. The details of the method for solving Eq.(\ref{sdg})
are outlined in the Appendix.

For the model parameters, we take $\alpha_s=0.368(3)$, $b=0.206(2)$
GeV$^2$, $\mu=0.056(11)$ GeV, $m_b=4.757(2)$ GeV, and $\sigma=3.10(25)$ GeV.
This parameter set is slightly different from that suggested in
Ref.~\cite{Chao:2009}. In our calculation, the cutoff distance
$r_c=0.060(12)$ fm is adopted. The uncertainties for these determined parameters
mean that if one changes one of the parameter within its uncertainty,
the mass change of one state is less than 5 MeV. It should be mentioned that the masses of
the $^3P_0$ states are sensitive to the cutoff distance $r_c$.
Thus, in the present work we use the mass of $\chi_{b0}(1P)$ to determine
the cutoff distance $r_c$. With the determined cutoff distance $r_c=0.06$ fm,
the calculated masses of the other $^3P_0$ states are in good agreement with
the measurements and the other model predictions.

With the determined parameter set, by solving the radial Schr\"{o}dinger
equation we obtain the masses of the bottomonium states, which have been listed in
Tab~\ref{tabs}. From the table, we see that our results are
compatible with the previous screened potential model
predictions~\cite{Chao:2009}, which indicates that our numerical
method is reliable. The recent relativized quark model can
successfully describe the low-lying bottomonium states, however,
their predicted mass for the higher excitations
$\Upsilon(6S)$ is about $100$ MeV larger than the
experimental measurements~\cite{Godfrey:2015dia}. Although the
recent nonrelativistic constituent quark model systematically
improve the descriptions of the higher mass spectrum, the
predicted masses for the ground states $\Upsilon(1S)$ and
$\eta_b(1S)$ are about $40\sim 50$ MeV higher than the
data~\cite{Segovia:2016xqb}. Interestedly, it is found that the
screened potential model obtains a fairly good description of the
masses not only for the low-lying states, but also for the higher
excitation $\Upsilon(6S)$.

Furthermore, in Tab.~\ref{tabhf}, we give our predictions of the
hyperfine splittings for some $S$-wave states, and fine splittings
for some $P$-wave states. It is found that our predicted splittings are in
good agreement with the world average data~\cite{PDG}.
Comparing the model predictions~\cite{Chao:2009,Godfrey:2015dia,
Segovia:2016xqb,Barducci:2016wze,Wei-Zhao:2013sta} with each other,
we find obvious model dependencies of the predicted mass splittings.
Thus, to better understand these nonperturbative strong interactions
in the bottomonium system, more model-independent studies are needed.


In order to clearly see the properties of the wave functions, we plot
the radial probability density of the states as a function of the interquark
distance $r$ in Fig.~\ref{wf}. It is found that the spin-dependent potentials have
notable corrections to the $S$- and triplet $P$-wave states; however,
the corrections to the triplet $D$-wave states are tiny. The strong
attractive spin-spin potential $H_{SS}$ shifts the wave functions of
the $^1S_0$ states towards the center, while the strong attractive
tensor potential $H_{T}$ shifts the wave functions of the $^3P_{0,1}$
states towards the center.

\section{electromagnetic transitions}\label{EMT}

Using these obtained wave functions of the bottomonium states, we further study
their EM transitions. The quark-photon EM coupling at the tree level
is adopted as
\begin{eqnarray}\label{he}
H_e=-\sum_j
e_{j}\bar{\psi}_j\gamma^{j}_{\mu}A^{\mu}(\mathbf{k},\mathbf{r})\psi_j,
\end{eqnarray}
where $\psi_j$ stands for the $j$th quark field in a hadron. The
photon has three momentum $\mathbf{k}$, and the constituent quark
$\psi_j$ carries a charge $e_j$.

To match the nonrelativistic wave functions of the bottomonium
states, we should adopt the nonrelativistic form of Eq.~(\ref{he})
in the calculations. For the EM transition of a hadron, in the
initial-hadron-rest system the nonrelativistic expansion of $H_e$ in
Eq.(\ref{he}) becomes~\cite{Li:1997gda,Deng:2015bva}
  \begin{equation}\label{he2}
h_{e}\simeq\sum_{j}\left[e_{j}\mathbf{r}_{j}\cdot\veps-\frac{e_{j}}{2m_{j}
}\vsig_{j}\cdot(\veps\times\hat{\mathbf{k}})\right]\phi,
\end{equation}
where $m_j$, $\vsig_j$, and $\mathbf{r}_j$ stand for the constituent mass,
Pauli spin vector, and coordinate for the $j$th quark, respectively.
The vector $\veps$ is the polarization vector of the photon.
For emitting a photon, we have $\phi=e^{-i\mathbf{k}\cdot
\mathbf{r}_{j}}$, while for absorbing a photon, we have $\phi=e^{+i\mathbf{k}\cdot
\mathbf{r}_{j}}$. It is found that the first and second terms in Eq.(\ref{he2}) are
responsible for the electric and magnetic transitions, respectively.
The main feature of this EM transition operator is that the effects
of binding potential between quarks are considered. Furthermore, the
possible higher EM multipole contributions are included naturally.
This nonrelativistic form has been widely applied to
meson photoproduction reactions~\cite{Li:1995si,Li:1997gda,Li:1994cy,
Zhao:2001kk, Saghai:2001yd, Zhao:2002id, He:2008ty, He:2008uf,
He:2010ii, Zhong:2011ti, Zhong:2011ht, Xiao:2015gra}. It should
be mentioned that, at the order of $1/m_j$, we have neglected the contributions from the
term $e_{j}\mathbf{r}_{j}\cdot\veps \mathbf{p}_j\cdot \mathbf{\hat{k}}/m_j$
as suggested in Refs.~\cite{Li:1994cy,Li:1995si} for a strong suppression of $\mathbf{p}_j\cdot \mathbf{\hat{k}}/m_j$.

Then, one obtains the standard helicity amplitude $\mathcal{A}$
of the radiative decay process by the relation
\begin{eqnarray}\label{amp3}
\mathcal{A}&=&-i\sqrt{\frac{\omega_\gamma}{2}}\langle f | h_{e}| i
\rangle.
\end{eqnarray}

Finally, we can calculate the EM decay width by
\begin{equation}\label{dww}
\Gamma=\frac{|\mathbf{k}|^2}{\pi}\frac{2}{2J_i+1}\frac{M_{f}}{M_{i}}\sum_{J_{fz},J_{iz}}|\mathcal{A}_{J_{fz},J_{iz}}|^2,
\end{equation}
where $J_i$ is the total angular momentum of an initial meson and
$J_{fz}$ and $J_{iz}$ are the components of the total angular
momenta along the $z$ axis of initial and final mesons,
respectively. In our calculation, for the well-established
bottomonium states, the experimental
masses are adopted~\cite{PDG}; while for the missing
bottomonium states, their masses are adopted from our theoretical
predictions.

\begin{table*}[htb]
\caption{ Partial widths of the M1 radiative transitions for some
low-lying $S$- and $P$-wave bottomonium states. For comparison, the
measured values from the PDG~\cite{PDG}, and the theoretical
predictions with the relativistic quark model~\cite{Ebert:2003}, nonrelativistic
effective field theories of
QCD (EFT model)~\cite{Pineda:2013lta}, relativized quark model (GI model)~\cite{Godfrey:2015dia}, and  nonrelativistic
constituent quark model (NR model)~\cite{Segovia:2016xqb} are also listed in
the same table. }\label{tab3}
\begin{tabular}{c|c|ccc|ccccc|cccc}  \hline\hline
 Initial meson       & Final meson & \multicolumn{3}{|c|} {\underline{$E_{\gamma}$  (MeV)}  } & \multicolumn{5}{|c|} {\underline{$\Gamma_{\mathrm{M1}}$  (eV)}} &\multicolumn{1}{|c|}{\underline{$\Gamma_{\mathrm{M1}}$  (eV)}}  \\
   state             & state            & Ref.\cite{Ebert:2003}& GI~\cite{Godfrey:2015dia}&  ours &   Ref.\cite{Ebert:2003}& GI~\cite{Godfrey:2015dia}& EFT~\cite{Pineda:2013lta} &NR~\cite{Segovia:2016xqb} & Ours &  PDG~\cite{PDG} \\
\hline
$\Upsilon(1 ^3S_{1})$      &$\eta_{b}(1 ^1S_{0})$       &  60  &  62 &  62  & $5.8$ & $10$  & $15.2$         &9.34 & $10$    &  \\
\hline
$\Upsilon(2 ^3S_{1})$      &$\eta_{b}(2 ^1S_{0})$       & 33   &  24 &  24  & $1.4$ & $0.59$& $0.67$         &0.58 & $0.59$  &  \\
                           &$\eta_{b}(1 ^1S_{0})$       & 604  & 606 & 606  & $6.4$ & $81$  & $6^{+26}_{-6}$ &56.5 &$66$     &$12.5\pm 4.9$\\
$\eta_{b}(2 ^1S_{0})$      &$\Upsilon(1 ^3S_{1})$       & 516  & 524 & 524  & $12$  & $68$  & $\sim 80$      &45.0 &$64$     & \\
\hline
$\Upsilon(3 ^3S_{1})$      &$\eta_{b}(3 ^1S_{0})$       & 27   &  18 &  18  & $0.8$ & $0.25$&                &0.66 &$3.9$   & \\
                           &$\eta_{b}(2 ^1S_{0})$       & 359  &  350&  350 & $1.5$ &  0.19 &                &11.0 &$11$    &$<14$\\
                           &$\eta_{b}(1 ^1S_{0})$       & 911  &  913&  913 & $11$  & $60$  &                &57.0 &$71$    &$10\pm 2$ \\
$\eta_{b}(3 ^1S_{0})$      &$\Upsilon(2 ^3S_{1})$       & 301  &  309&  309 & $2.8$ & $9.1$ &                &9.20 &$8.7$    &  \\
                           &$\Upsilon(1 ^3S_{1})$       & 831  &  840&  840 & $24$  & $74$  &                &51.0 &$60$    &  \\
\hline
$\chi_{b2}(1 ^3P_{2})$     &$h_{b}(1 ^1P_{1})$          &      &  13 &  13  &       & $9.6\times10^{-2}$& $0.12$  &$8.9\times10^{-2}$  & $9.5\times10^{-2}$    & \\
$h_{b}(1 ^1P_{1})$         &$\chi_{b1}(1 ^3P_{1})$      &      &  6  &  6   &       & $1.0\times10^{-2}$& $9.0\times10^{-3}$ &$1.15\times10^{-2}$   & $9.4\times10^{-3}$   & \\
                           &$\chi_{b0}(1 ^3P_{0})$      &      &  40 &  40  &       & $0.89$&  0.96        &0.86 & $0.90$   & \\
\hline
$\chi_{b2}(2 ^3P_{2})$     &$h_{b}(1 ^1P_{1})$          &      &  363&  363 &       & $0.24$&   &1.78  & $4.5$   &  \\
$\chi_{b1}(2 ^3P_{1})$     &                            &      &  350&  350 &       & $2.2$&    &0.17 & $0.18$   &  \\
$\chi_{b0}(2 ^3P_{0})$     &                            &      &  329&  329 &       & $9.7$&    &2.39 & $16$   &  \\
$h_{b}(2 ^1P_{1})$         &$\chi_{b2}(1 ^3P_{2})$      &      &  342&  342 &       & $2.2$&    &$6.91\times10^{-3}$ & $1.1$   &  \\
                           &$\chi_{b1}(1 ^3P_{1})$      &      &  360&  360 &       & $1.1$&    &1.28 & $2.5$   &  \\
                           &$\chi_{b0}(1 ^3P_{0})$      &      &  393&  393 &       & $0.32$&   &36.4  & $10$   &  \\
\hline\hline
\end{tabular}
\end{table*}


\subsection{$\Upsilon (1S)\rightarrow \eta_b(1S) \gamma$}

The $\Upsilon (1S)\rightarrow \eta_b(1S) \gamma$ decay process is a
typical M1 transition at tree level, which is strongly suppressed by
the constituent bottom quark mass $m_{b}$. Our predicted partial
width is
\begin{eqnarray}
\Gamma[\Upsilon (1S)\rightarrow \eta_b(1S) \gamma]& \simeq &
10 \ \mathrm{eV}.
\end{eqnarray}
Combining this partial width with the measured total width of $\Upsilon (1S)$~\cite{PDG},
we obtain
\begin{eqnarray}
\mathcal{B}[\Upsilon (1S)\rightarrow \eta_b(1S) \gamma]\simeq
2.0\times 10^{-4}.
\end{eqnarray}
Our predictions are in good agreement with the recent results of the
relativized quark model ~\cite{Godfrey:2015dia} and nonrelativistic
constituent quark model~\cite{Segovia:2016xqb} (see
Tab.~\ref{tab3}). However, our predicted $\Gamma[\Upsilon
(1S)\rightarrow \eta_b(1S) \gamma]$ is larger than the value $5.8$
eV from the relativistic quark model~\cite{Ebert:2003}, while
smaller than the recent prediction $15.2$ eV from the pNRQCD
approach~\cite{Pineda:2013lta}. It should be mentioned that this
decay rate is extremely sensitive to the masses of $\Upsilon (1S)$
and $\eta_b(1S)$. If all of the models adopt the experimental
masses, the predictions of $\Gamma[\Upsilon (1S)\rightarrow
\eta_b(1S) \gamma]$ from different models might be consistent with
each other.

\subsection{Radiative transitions of $2S$ states}

\subsubsection{$\Upsilon(2S)$}

The allowed EM transitions of $\Upsilon(2S)$ are $\Upsilon(2S)\to
\chi_{bJ}(1P) \gamma$ and $\Upsilon(2S)\to \eta_b(1S,2S)\ \gamma$.
The $\Upsilon(2S)\to \chi_{bJ}(1P) \gamma$ processes are
governed by the E1 transitions, while $\Upsilon(2S)\to
\eta_b(1S,2S)\ \gamma$ are typical M1 transitions.

From Tab.~\ref{tab1}, it is found that our predicted partial decay widths for the $\Upsilon(2S)\to
\chi_{bJ}(1P)\gamma$ processes are in good agreement with the world average
data from the PDG~\cite{PDG}, and also are consistent with predictions
from various potential models~\cite{Chao:2009,Ebert:2003,Barducci:2016wze,
Wei-Zhao:2013sta,Godfrey:2015dia,Segovia:2016xqb}.

For the M1 transitions $\Upsilon(2S)\to
\eta_b(1S,2S)\gamma$, our predicted partial decay
widths have been listed in Tab.~\ref{tab3}. From the table, it is seen that
our predicted $\Gamma[\Upsilon(2S)\to\eta_b(2S)\gamma]$ is in agreement with the other model predictions.
It should be pointed out that although our predicted $\Gamma[\Upsilon(2S)\to \eta_{b}(1S)\gamma]$ is compatible
with the recent potential model predictions~\cite{Godfrey:2015dia,Segovia:2016xqb}, it is
about $5$ times larger than the average value  $1.25(49)\times
10^{-2} \ \mathrm{keV}$ from the PDG~\cite{PDG} and the recent lattice NRQCD
result $1.72(55)\times 10^{-2} \
\mathrm{keV}$~\cite{Hughes:2015dba}.
More studies of the M1 transition $\Upsilon(2S)\to
\eta_b(1S)\gamma$ are needed in both theory and experiments.

\subsubsection{$\eta_b(2S)$}

The $\eta_b(2S)$ resonance can decay into $h_b(1P)\gamma$ and
$\Upsilon(1S) \gamma$ channels by the E1 and M1 transitions, respectively.
Our predicted partial decay width of $\Gamma[\eta_b(2S)\to h_b(1P)\gamma] \simeq  3.41$ keV
is in good agreement with the predictions of the relativistic
quark model~\cite{Ebert:2003}, potential
model~\cite{Godfrey:2015dia}, and nonrelativistic constituent quark model~\cite{Segovia:2016xqb}
(see Tab.~\ref{tab1}). However, it is about a factor 1.8 smaller
than the previous SNR model prediction~\cite{Chao:2009}. This difference might come from
the corrections of the spin-dependent potentials to the wave function of $\eta_b(2S)$.

Furthermore, from Tab.~\ref{tab3} it is seen that our predicted partial decay width for the M1 transition
$\eta_b(2S)\to \Upsilon(1S) \gamma$
is compatible with the recent predictions of
potential models~\cite{Godfrey:2015dia,Segovia:2016xqb}, and with the pNRQCD approach~\cite{Pineda:2013lta}.
However, our prediction is notably larger than $\Gamma[\eta_b(2S)\to \Upsilon(1S) \gamma]\simeq 12$ eV predicted
with a relativistic quark model~\cite{Ebert:2003} (see Tab.~\ref{tab3}).

\subsection{Radiative transitions of $1P$ states}


The typical radiative transitions of $\chi_{bJ}(1P)$ are
$\chi_{bJ}(1P)\to \Upsilon(1S) \gamma$.
From the Tab.~\ref{tab2}, it is found that the partial widths
$\Gamma[\chi_{bJ}(1P)\to \Upsilon(1S) \gamma]$ predicted by us
are in agreement with the predictions in~\cite{Chao:2009,Ebert:2003,
Wei-Zhao:2013sta,Godfrey:2015dia,Segovia:2016xqb}. Combining our predicted
partial widths with the measured branching ratios
$\mathcal{B}[\chi_{b0}(1P)\to \Upsilon(1S)\gamma]\simeq (1.76\pm 0.48)\%$,
$\mathcal{B}[\chi_{b1}(1P)\to \Upsilon(1S)\gamma]\simeq (33.9\pm 2.2)\%$, and
$\mathcal{B}[\chi_{b2}(1P)\to \Upsilon(1S)\gamma]\simeq
(19.1\pm 1.2)\%$~\cite{PDG}, we easily estimate the total widths
for $\chi_{b0}(1P)$, $\chi_{b1}(1P)$ and $\chi_{b2}(1P)$, which are
\begin{eqnarray}
\Gamma^{\mathrm{total}}_{\chi_{b0}(1P)}&\simeq & 1.56^{+0.59}_{-0.33} \ \mathrm{MeV},\\
\Gamma^{\mathrm{total}}_{\chi_{b1}(1P)}&\simeq & 94\pm 7\ \mathrm{keV},\\
\Gamma^{\mathrm{total}}_{\chi_{b2}(1P)}&\simeq & 166^{+12}_{-9}\ \mathrm{keV},
\end{eqnarray}
respectively. It is interesting to find that the estimated width for $\chi_{b0}(1P)$ is consistent with the
recent measurement $1.3\pm 0.9$ MeV from the Belle Collaboration~\cite{Abdesselam:2016xbr}.
It should be mentioned that these widths predicted by us
strongly depend on the measured branching ratios. It is found that
$\mathcal{B}[\chi_{b0}(1P)\to \Upsilon(1S)\gamma]$ still bears a large uncertainty.
Thus, to determine finally the width of $\chi_{b0}(1P)$, more accurate measurements are needed.

For the singlet state $h_{b}(1P)$, its main radiative
transition is $h_{b}(1P)\to \eta_b(1S) \gamma$.  We predict that
$\Gamma[h_{b}(1P)\to \eta_{b}(1S) \gamma]\simeq35.8 \
\mathrm{keV}$, which is consistent with the
predictions in Refs.~\cite{Chao:2009,Godfrey:2015dia} (see
Tab.~\ref{tab2}). A relatively large partial width was also predicted in
Ref~\cite{Ebert:2003}. Combining the measured branching ratio
$\mathcal{B}[h_{b}(1P)\to \eta_{b} \gamma]\simeq
49^{+8}_{-7}\%$~\cite{PDG} with our predicted partial width, we
estimate that the total width of $h_{b}(1P)$ might be
\begin{eqnarray}
\Gamma^{\mathrm{total}}_{h_{b}(1P)}\simeq 73^{+12}_{-10}\ \ \mathrm{keV},
\end{eqnarray}
which could be tested in future experiments.

Finally, we give our estimations of the typical M1 transitions
$h_{b}(1P)\to \chi_{b0,1}(1P) \gamma$, which are listed in Tab.~\ref{tab3}.
The rates of these M1 transitions are very weak.
Our results are consistent with those obtained in the framework of the relativized quark model~\cite{Godfrey:2015dia}
and nonrelativistic constituent quark model~\cite{Segovia:2016xqb}.

\subsection{Radiative transitions of $1D$ states}

In the $1D$ bottomonium states, only the $2^{--}$ state
$\Upsilon_2(1D)$ with a mass of $M_{\Upsilon_2(1D)}=10164$ MeV is
confirmed in experiments~\cite{delAmoSanchez:2010kz}. The other $1D$
states are still missing. The discovery
of the $\Upsilon_2(1D)$ state provides a strong constrain on the
masses of the other $1D$ states. In our calculations,
we predict the mass splittings $M_{\Upsilon_3(1D)}-M_{\Upsilon_2(1D)}\simeq 4$,
$M_{\Upsilon_2(1D)}-M_{\Upsilon_1(1D)}\simeq 7$, and
$M_{\Upsilon_2(1D)}-M_{\eta_{b2}(1D)}\simeq 0$ MeV.
Combining these predicted multiplet
mass splittings with the measured mass of $M_{\Upsilon_2(1D)}=10164$
MeV, one can predict the masses for the $\Upsilon_1(1D)$,
$\Upsilon_3(1D)$ and $\eta_{b2}(1D)$ states, which are
$M_{\Upsilon_1(1D)}\simeq 10157$ MeV, $M_{\Upsilon_3(1D)}\simeq 10168$ MeV, and
$M_{\eta_{b2}(1D)}\simeq 10164$ MeV, respectively.

\subsubsection{$\Upsilon_2(1D)$}

For the established  $2^{--}$ state $\Upsilon_2(1D)$ [i.e.,
$\Upsilon_2(10164)$], the EM transitions are dominated by
$\Upsilon_2(1D)\to \chi_{b1,2}(1P)\gamma$. We calculate their partial decay widths, which
are listed in Tab.~\ref{tab2}. Combining with
the predicted partial widths of $\Gamma[\Upsilon_2(1D)\to ggg]\simeq0.62$ keV
and $\Gamma[\Upsilon_2(1D)\to\pi\pi \Upsilon(1S)]\simeq0.29$ keV from Ref.~\cite{Segovia:2016xqb},
we estimate the total width of $\Upsilon_2(1D)$,
$\Gamma_{\mathrm{tot}}\simeq 30$ keV. With this estimated width,
we further predict the branching ratios
\begin{eqnarray}
\mathcal{B}[\Upsilon_2(1D)\to \chi_{b1}(1P) \gamma]& \simeq & 73\%,\\
\mathcal{B}[\Upsilon_2(1D)\to \chi_{b2}(1P) \gamma]& \simeq & 24\% .
\end{eqnarray}
Our results are in agreement with the predictions obtained with the previous SNR
model~\cite{Chao:2009}, relativistic quark
model~\cite{Ebert:2003}, and nonrelativistic constituent quark model
~\cite{Segovia:2016xqb}. The large branching ratios indicate the
$\Upsilon_2(1D)\to \chi_{b1,2}(1P)\gamma$ transitions may be
observed in forthcoming experiments.

\subsubsection{The missing $1D$ states}

According to the predicted mass $M_{\Upsilon_1(1D)}=10157$ MeV of
$\Upsilon_1(1D)$, we calculate the partial decay widths of
$\Gamma[\Upsilon_1(1D)\to \chi_{b0,1,2}(1P)\gamma]$, which are listed
in Tab.~\ref{tab2}. In Ref.~\cite{Segovia:2016xqb}, the total width of $\Upsilon_1(1D)$
is predicted to be $\Gamma_{\mathrm{tot}}\simeq 44$ keV. Using it as an input, we predict
\begin{eqnarray}
\mathcal{B}[\Upsilon_1(1D)\to \chi_{b0}(1P) \gamma]& \simeq & 45\%,\\
\mathcal{B}[\Upsilon_1(1D)\to \chi_{b1}(1P) \gamma]& \simeq & 30\%,\\
\mathcal{B}[\Upsilon_1(1D)\to \chi_{b2}(1P) \gamma]& \simeq & 2\% .
\end{eqnarray}
These branching ratios are consistent with those from the recent works~\cite{Godfrey:2015dia,Segovia:2016xqb}.
The fairly large branching ratios indicate that the missing $\Upsilon_1(1D)$ state
is most likely to be observed through the radiative transitions $\Upsilon_1(1D)\to \chi_{b0,1}(1P)\gamma$.

While taking the mass of $\Upsilon_3(1D)$ with
$M_{\Upsilon_3(1D)}=10168$ MeV, we calculate the partial decay widths
of $\Gamma[\Upsilon_3(1^3D_3)\to \chi_{bJ}(1P)\gamma]$. Our
results are listed in Tab.~\ref{tab2}.
It is found that the EM decays of $\Upsilon_3(1D)$ are governed by
the $\chi_{b2}(1P)\gamma$ channel, and the decay rates into the
$\chi_{b0,1}(1P)\gamma$ channels are negligibly small. Our
prediction of $\Gamma[\Upsilon_3(1D)\to \chi_{b2}(1P) \gamma] \simeq
32.1$ keV is consistent with the predictions from the potential
models~\cite{Chao:2009,Godfrey:2015dia} and relativistic quark
model~\cite{Ebert:2003} (see Tab.~\ref{tab2}). According to
the predictions in Refs.~\cite{Godfrey:2015dia,Segovia:2016xqb},
the partial widths of $\Gamma[\Upsilon_2(1D)\to ggg]$
and $\Gamma[\Upsilon_2(1D)\to\pi\pi \Upsilon(1S)]$ are
too small to compare with $\Gamma[\Upsilon_3(1D)\to \chi_{b2}(1P) \gamma]$, thus,
the branching fraction of $\mathcal{B}[\Upsilon_3(1D)\to \chi_{b2}(1P) \gamma]\sim 100\%$.
To establish $\Upsilon_3(1D)$, the decay channel $\chi_{b2}(1P) \gamma$ is worth observing in future experiments.

For the singlet $1D$ state $\eta_{b2}(1D)$, our predicted partial width
$\Gamma[\eta_{b2}(1D)\to  h_{b}(1P) \gamma] \simeq  30.3$ keV
is close to the predictions from the other potential
models~\cite{Chao:2009,Ebert:2003,Godfrey:2015dia} (see
Tab.~\ref{tab2}). Combining with the predictions $\Gamma[\eta_{b2}(1D)\to gg]\simeq 1.8$ keV
and $\Gamma[\eta_{b2}(1D)\to\pi\pi \eta_b(1S)]\simeq 0.35$ keV
in Ref.~\cite{Godfrey:2015dia}, we obtain the total width of $\Upsilon_1(1D)$,
$\Gamma_{\mathrm{tot}}\simeq 32.5$ keV, with which we further estimate that
\begin{eqnarray}
\mathcal{B}[\eta_{b2}(1D)\to  h_{b}(1P) \gamma]& \simeq & 93\%.
\end{eqnarray}
The large radiative transition rate indicates
that the missing $\eta_{b2}(1D)$ state is most likely to be observed in
the $h_{b}(1P) \gamma$ channel.

\subsection{Radiative transitions of $2P$ states}

The $2P$ bottomonium states have been established in
experiments. The branching ratios of
$\mathcal{B}[\chi_{b0,1,2}(2P)\to\Upsilon(1S,2S) \gamma]$ and
$\mathcal{B}[h_b(2P)\to\eta_b(1S,2S) \gamma]$ have been measured. These measured
branching ratios give us a good chance to study the radiative
transitions of the $2P$ bottomonium states, and test our model.

\subsubsection{$\chi_{b0}(2P)$}

The allowed EM decay modes of $\chi_{b0}(2P)$ are $\Upsilon(1S,2S)
\gamma$, $\Upsilon_1(1D)\gamma$ and $h_b(1P)\gamma$. We calculate
their partial widths and list them in Tab.~\ref{tab1}. From the table,
one can see that our predictions are compatible with the other model predictions.
Taking the predicted total width $\Gamma_{\mathrm{tot}}\simeq 2.5$ MeV of $\chi_{b0}(2P)$ from
Ref.~\cite{Godfrey:2015dia} as an input, we further predict that
\begin{eqnarray}
\mathcal{B}[\chi_{b0}(2P)\to \Upsilon(1S) \gamma]& \simeq & 2.2\times 10^{-3},\\
\mathcal{B}[\chi_{b0}(2P)\to \Upsilon(2S) \gamma]& \simeq & 5.8\times 10^{-3}.
\end{eqnarray}
Our prediction is compatible with the recent results obtained from potential models
~\cite{Godfrey:2015dia,Segovia:2016xqb}, and the previous results obtained from SNR$_1$ model~\cite{Chao:2009}.
However, the predicted branching ratio $\mathcal{B}[\chi_{b0}(2P)\to \Upsilon(2S) \gamma]$ is about
an order of magnitude smaller than the data from the PDG~\cite{PDG}.
To test our predictions, more accurate measurements are needed in experiments.

We also study the typical M1 transition $\chi_{b0}(2P)\to
h_b(1P) \gamma$. Our predicted partial decay width
$\Gamma[\chi_{b0}(2P)\to h_b(1P) \gamma] \simeq 1.6\times 10^{-2}$ keV
is close to the recent predictions with the GI potential model~\cite{Godfrey:2015dia} (see
Tab.~\ref{tab3}).


\subsubsection{$\chi_{b1}(2P)$}

The $\chi_{b1}(2P)$ state can decay into $\Upsilon(1S,2S)\gamma$,
$\Upsilon(1^3D_{2,3})\gamma$ and $h_b(1P)\gamma$ via radiative
transitions. Our predicted partial widths for these transitions
are listed in Tab.~\ref{tab1}. From the table it is found that
the decay rates of $\chi_{b1}(2P)$ into the $D$-wave states
$\Upsilon_{1,2}(1D)$ are much weaker than those into the $S$-wave
states. Our predicted partial widths of $\Gamma[\chi_{b1}(2P)\to \Upsilon(1S,2S) \gamma]$
are consistent with observations from the CLEO Collaboration~\cite{Cawlfield:2005ra}.
Combining our predicted partial widths with the total width $\Gamma_{\mathrm{tot}}\simeq 133$ keV
predicted in Ref.~\cite{Segovia:2016xqb}, we obtain that
\begin{eqnarray}
\mathcal{B}[\chi_{b1}(2P)\to \Upsilon(1S) \gamma]& \simeq & 8.1\%,\\
\mathcal{B}[\chi_{b1}(2P)\to \Upsilon(2S) \gamma]& \simeq & 11.5\%,
\end{eqnarray}
which are close to the measured values $\mathcal{B}[\chi_{b1}(2P)\to \Upsilon(1S) \gamma] \simeq  9.2\pm 0.8\%$ and
$\mathcal{B}[\chi_{b1}(2P)\to \Upsilon(2S) \gamma] \simeq  19.9\pm 1.9\%$~\cite{PDG}.
The branching fraction ratio
\begin{eqnarray}
\frac{\Gamma[\chi_{b1}(2P)\to \Upsilon(2S)
\gamma]}{\Gamma[\chi_{b1}(2P)\to \Upsilon(1S) \gamma]}& \simeq &
1.4,
\end{eqnarray}
is slightly smaller than the world average value $2.2\pm 0.4$
from the PDG~\cite{PDG}. From Tab.~\ref{tab1}, we can find that this
ratio has a strong model dependency. To test the predictions from
various models, more accurate measurements are needed in experiments.

Furthermore, the typical M1 transition $\chi_{b2}(2P)\to h_b(1P) \gamma$ is also
studied. The predicted partial decay width
\begin{eqnarray}
\Gamma[\chi_{b2}(2P)\to h_b(1P) \gamma]& \simeq & 1.8\times 10^{-4}
\ \mathrm{keV},
\end{eqnarray}
is about an order of magnitude smaller than the recent prediction
$2.2\times 10^{-3}$ keV in Ref.~\cite{Godfrey:2015dia}.
However, the recent prediction $1.7\times 10^{-4}$ keV with a nonrelativistic constituent
quark model~\cite{Segovia:2016xqb} is in good agreement with our prediction.
The Lattice QCD study may be able to clarify this puzzle.

\subsubsection{$\chi_{b2}(2P)$}

The $\chi_{b2}(2P)$ state can decay into $\Upsilon(1S,2S)\gamma$,
$\Upsilon_{1,2,3}(1D)\gamma$ and $h_b(1P)\gamma$ channels.
In these decays, the $\chi_{b2}(2P)\to
\Upsilon(1S,2S)\gamma$ processes play dominant roles. From Tab.~\ref{tab1}, it is seen that
our predicted partial widths of $\Gamma[\chi_{b2}(2P)\to \Upsilon(1S,2S) \gamma]$
are compatible with the observations from the CLEO Collaboration~\cite{Cawlfield:2005ra}
and other model predictions~\cite{Ebert:2003,Chao:2009,
Godfrey:2015dia,Segovia:2016xqb}. Combining our predicted partial widths with the estimated total width of $\chi_{b2}(2P)$ according
to the CLEO observations~\cite{Cawlfield:2005ra}, i.e.,
$\Gamma_{\mathrm{tot}}\simeq 143$ keV, we have
\begin{eqnarray}
\mathcal{B}[\chi_{b2}(2P)\to \Upsilon(1S) \gamma]& \simeq & 9.5\%,\\
\mathcal{B}[\chi_{b2}(2P)\to \Upsilon(2S) \gamma]& \simeq & 11\%,
\end{eqnarray}
which are close to the average data from the PDG~\cite{PDG}.
The estimated partial width ratio
\begin{eqnarray}
\frac{\Gamma[\chi_{b2}(2P)\to \Upsilon(2S)
\gamma]}{\Gamma[\chi_{b2}(2P)\to \Upsilon(1S) \gamma]}& \simeq &
1.2,
\end{eqnarray}
is also close to the lower limit of the world average data
$1.51_{-0.47}^{+0.59}$ from the PDG~\cite{PDG}. This ratio
has strong model dependencies. Thus, more accurate measurements are
needed to test various model predictions.

The decay rates of $\chi_{b2}(2P)\to \Upsilon_{1,2,3}(1D)\gamma$ are
much weaker than those of $\chi_{b2}(2P)\to \Upsilon(1S,2S)\gamma$.
Our predicted results are close to the predictions in
Refs.~\cite{Ebert:2003,Chao:2009,Segovia:2016xqb} (see Tab.~\ref{tab1}).
Combining the estimated total width of $\chi_{b2}(2P)$ with our predicted partial widths, we have
\begin{eqnarray}
\mathcal{B}[\chi_{b2}(2P)\to  \Upsilon_1(1D)  \gamma]& \simeq & 1.8\times 10^{-4},\\
\mathcal{B}[\chi_{b2}(2P)\to  \Upsilon_2(1D)  \gamma]& \simeq & 2.9\times 10^{-3},\\
\mathcal{B}[\chi_{b2}(2P)\to  \Upsilon_3(1D)  \gamma]& \simeq & 1.7\times 10^{-2}.
\end{eqnarray}
To look for the missing $\Upsilon_3(1D)$ state, the
three-photon decay chain $\chi_{b2}(2P)\to \Upsilon_3(1D) \gamma \to \chi_{b2}(1P)\gamma\gamma
\to \Upsilon(1S)\gamma\gamma\gamma$ is worth observing.
The combined branching ratio can reach up to $\mathcal{O}(10^{-3})$.

\subsubsection{$h_{b}(2P)$}

The $h_{b}(2P)$ state can decay into $\eta_b(1S,2S)\gamma$,
$\eta_{b2}(1D)\gamma$, and $\chi_{b0,1,2}(1P)\gamma$ via EM
transitions, in which the $\eta_b(1S,2S)\gamma$ decay modes are
dominant. We calculate the partial decay widths of $\Gamma[h_{b}(2P)\to
\eta_b(1S,2S)\gamma]$, which are listed in Tab.~\ref{tab1}. Our results are
compatible with the other
model predictions~\cite{Ebert:2003,Chao:2009,
Godfrey:2015dia,Segovia:2016xqb}. Our predicted partial width ratio,
\begin{eqnarray}
\frac{\Gamma[h_{b}(2P)\to \eta_b(2S) \gamma]}{\Gamma[h_{b}(2P)\to
\eta_b(1S) \gamma]}& \simeq & 1.0,
\end{eqnarray}
is close to the lower limit of the measurement $1.0\pm 4.3$ from the Belle
Collaboration~\cite{Mizuk:2012pb}. Furthermore, combining the measured
branching ratio $\mathcal{B}[h_{b}(2P)\to \eta_b(1S) \gamma]\simeq
22.3\pm 3.8^{+3.1}_{-3.3}\%$ with our predicted partial width, we
estimate the total width of $h_{b}(1P)$, which is
\begin{eqnarray}\label{hb2p}
\Gamma^{\mathrm{total}}_{h_{b}(2P)}\simeq 72^{+34}_{-17}\ \ \mathrm{keV}.
\end{eqnarray}
It could be tested in future experiments.

We also study the transition of $h_{b}(2P)\to
\eta_{b2}(1D) \gamma$. The predicted partial width
$\Gamma[h_{b}(2P)\to \eta_{b2}(1D)  \gamma] \simeq  2.24$ keV
is compatible with the predictions from the
relativized quark model~\cite{Godfrey:2015dia} and the relativistic quark
model~\cite{Ebert:2003}. Using this predicted total width in Eq.~(\ref{hb2p}) as an input,
we further predict
\begin{eqnarray}
\mathcal{B}[h_{b}(2P)\to \eta_{b2}(1D)  \gamma]& \simeq & 3\%.
\end{eqnarray}
Combining this ratio with our predicted ratio of $\mathcal{B}[\eta_{b2}(1D)\to  h_{b}(1P) \gamma] \simeq  93\%$
and the measured ratios of $\mathcal{B}[h_b(1P)\to  \eta_b\gamma] \simeq  49\%$,
we obtain the combined branching ratio for the three-photon cascade $h_{b}(2P)\to\eta_{b2}(1D)  \gamma \to h_b(1P)\gamma\gamma\to \eta_b\gamma\gamma\gamma$:
\begin{eqnarray}
\mathcal{B}[h_{b}(2P)\to\eta_{b2}(1D)  \gamma \to h_b(1P)\gamma\gamma\to \eta_b\gamma\gamma\gamma] \simeq 1.4\%.
\end{eqnarray}
Thus, to establish the missing $\eta_{b2}(1D)$ this three-photon cascade is worth observing.

Finally, we give our predictions for the typical M1 transitions
$h_{b}(2P)\to \chi_{b0,1,2}(1P) \gamma$. Our results are listed in Tab.~\ref{tab3}. It is seen that
concerning these M1 transitions, there are obvious differences in various model predictions.

\subsection{Radiative transitions of $3S$ states}

\subsubsection{$\Upsilon(3S)$}

$\Upsilon(3S)$ is well established in experiments. Its mass and
width are $M_{\Upsilon(3S)}=10355.2\pm 0.5$ MeV and $\Gamma=20.32\pm
1.85$ keV, respectively. The EM transitions $\Upsilon(3S)\to
\chi_{bJ}(1P,2P)\gamma$ and $\Upsilon(3S)\to \eta_{b}(1S,2S)\gamma$
have been observed in experiments. We calculate the partial widths and
compare them with the data in Tab.~\ref{tab1}.

From the table, it is found that for the EM transitions $\Upsilon(3S)\to \chi_{bJ}(1P)\gamma$, the
predicted partial widths are in good agreement with the world average data
from the PDG~\cite{PDG}. Note that the transition widths for
$\Upsilon(3S)\to \chi_{b1,2}(1P)\gamma$ calculated from the previous
screened potential model~\cite{Chao:2009} are too large as compared with experimental data.
These problems have been overcome in our calculations by considering the
corrections of the spin-dependent interactions to the wave functions.
It indicates that the corrections of the spin-dependent
interactions to the wave functions are important to understand
these EM transitions, which was also found in Ref.~\cite{Badalian:2012mb}.

While for the EM transitions $\Upsilon(3S)\to \chi_{bJ}(2P)\gamma$,
from Tab.~\ref{tab1} it is found that our predicted partial widths of
$\Gamma[\Upsilon(3S)\to \chi_{bJ}(2P)\gamma]$ are in good agreement with the experimental data and the
predictions in Refs.~\cite{Chao:2009,Ebert:2003,Barducci:2016wze,
Wei-Zhao:2013sta,Godfrey:2015dia,Segovia:2016xqb}. Combining our predicted partial
widths with the measured width of $\Upsilon(3S)$, we estimate that
\begin{eqnarray}
\mathcal{B}[\Upsilon(3S)\to \chi_{b0}(2P) \gamma]& \simeq & 5.5\%,\\
\mathcal{B}[\Upsilon(3S)\to \chi_{b1}(2P) \gamma]& \simeq & 12.8\%,\\
\mathcal{B}[\Upsilon(3S)\to \chi_{b2}(2P) \gamma]& \simeq & 15.6\%,
\end{eqnarray}
which are also in good agreement with the data from the PDG~\cite{PDG}.

For the typical M1 transitions $\Upsilon(3S)\to
\eta_{b}(1S,2S)\gamma$, our predicted partial widths are listed in Tab.~\ref{tab3}.
Our results are the same order of magnitude as the predictions from the
recent nonrelativistic constituent quark model
~\cite{Segovia:2016xqb}. However, our prediction of the $\Gamma[\Upsilon(3S)\to \eta_{b}(1S) \gamma]
\simeq  71\ \mathrm{eV}$ is notably larger than
the world average data $10\pm 2$ eV~\cite{PDG}. To clarify this puzzle, more studies are needed.

\subsubsection{$\eta_b(3S)$}

The $3^1S_0$ state, $\eta_b(3S)$, is still missing. The
predicted mass splitting between $3^3S_1$ and $3^1S_0$ is about $17$
MeV. Combining it with the measured mass of $3^3S_1$, we predict that the mass of
$\eta_b(3S)$ might be $M_{\eta_b(3S)}\simeq 10338$ MeV. Using this
predicted mass, we study the E1 transitions $\eta_b(3S)\to
h_b(1P,2P)\gamma$ and M1 transitions $\eta_b(3S)\to
\Upsilon(1S,2S)\gamma$. Our results have been listed in
Tabs.~\ref{tab3} and ~\ref{tab1}.

From Tab.~\ref{tab1}, it is
found that with the corrections of the spin-dependent potentials
to the wave functions, our predicted partial widths for the E1 transitions
$\eta_b(3S)\to h_b(1P,2P)\gamma$  are about a factor 2 smaller
than the previous screened potential model predictions~\cite{Chao:2009}.
Furthermore, it should be mentioned that our predicted partial width ratio
\begin{eqnarray}
\frac{\Gamma[\eta_b(3S)\to h_b(2P)  \gamma]}{\Gamma[\eta_b(3S)\to
h_b(1P) \gamma]}&\simeq & 6.1,
\end{eqnarray}
is notably different from the other model predictions~\cite{Chao:2009,Ebert:2003,
Godfrey:2015dia,Segovia:2016xqb}. From Tab.~\ref{tab3}, it is
found that our predicted partial widths for the M1 transitions $\eta_b(3S)\to \Upsilon(1S,2S)\gamma$
are compatible with the recent predictions in Refs.
~\cite{Godfrey:2015dia,Segovia:2016xqb}, however, our predictions are about
a factor 3 larger than the predictions with the relativistic quark
model~\cite{Ebert:2003}. These radiative transitions should be further studied in theory.

\subsection{Radiative transitions of $2D$ states}

Until now, no $2D$ bottomonium states have been observed in
experiments. In our calculations, their masses are adopted from
our potential model predictions.

\subsubsection{$\Upsilon_3(2D)$}

The radiative transitions of $\Upsilon_3(2D)$ are dominated by the
$\chi_{b2}(2P)\gamma$ channel, and the partial width decaying into
the $\chi_{b2}(1P)\gamma$ channel is also sizeable. Taking the mass of
$M_{\Upsilon_3(2D)}=10436$ MeV predicted by us, we calculate
the partial widths of $\Gamma[\Upsilon_3(2D)\to \chi_{b2}(1P,2P) \gamma]$. The results
compared with the other model predictions are listed in Tab.~\ref{tab2},
where we can see that our predictions
are compatible with the other model predictions. In Ref.~\cite{Godfrey:2015dia},
the total width of $\Upsilon_3(2D)$ is predicted to be
$\Gamma_{\mathrm{tot}}\simeq 25$ keV. With this predicted width, we further estimate
the branching ratios:
\begin{eqnarray}
\mathcal{B}[\Upsilon_3(2D)\to \chi_{b2}(1P) \gamma]& \simeq & 21\%,\\
\mathcal{B}[\Upsilon_3(2D)\to \chi_{b2}(2P) \gamma]& \simeq & 68\%.
\end{eqnarray}
To establish the $\Upsilon_3(2D)$ state, the $\chi_{b2}(1P,2P) \gamma$ channels
are worth observing.

\subsubsection{$\Upsilon_2(2D)$}

The radiative transitions of $\Upsilon_2(2D)$ are dominated by the
$\chi_{b1}(2P)\gamma$ channel, and the partial widths decaying into
the $\chi_{b2}(2P)\gamma$, $\chi_{b1}(1P)\gamma$ and
$\chi_{b2}(1P)\gamma$ channels are also sizeable. With the predicted
mass $M_{\Upsilon_2(2D)}=10432$ MeV, we predict
the partial widths for these radiative transitions. Our results compared with
the other model predictions are listed in Tab.~\ref{tab2}. From the table,
it is seen that the partial widths predicted by us are comparable with the
other model predictions in magnitude~\cite{Chao:2009,Ebert:2003,
Godfrey:2015dia,Segovia:2016xqb}. However, it should be mentioned that the predicted ratios
from different models are very different. In Ref.~\cite{Godfrey:2015dia},
the total width of $\Upsilon_2(2D)$ is predicted to be
$\Gamma_{\mathrm{tot}}\simeq 23$ keV. With this predicted total width, we further estimate that
\begin{eqnarray}
\mathcal{B}[\Upsilon_2(2D)\to \chi_{b1}(2P) \gamma]& \simeq & 50\%,\\
\mathcal{B}[\Upsilon_2(2D)\to \chi_{b2}(2P) \gamma]& \simeq & 16\%,\\
\mathcal{B}[\Upsilon_2(2D)\to \chi_{b1}(1P) \gamma]& \simeq & 17\%,\\
\mathcal{B}[\Upsilon_2(2D)\to \chi_{b2}(1P) \gamma]& \simeq & 5\%.
\end{eqnarray}
Observation of the $\chi_{b1,2}(2P)\gamma$ and $\chi_{b1}(1P)\gamma$
channels may be crucial to establish the missing $\Upsilon_2(2D)$ state.

\subsubsection{$\Upsilon_1(2D)$}

The radiative transitions of $\Upsilon_1(2D)$ are dominated by the $\chi_{b0,1}(2P)\gamma$
channels, and the partial widths decaying into the $\chi_{b0,1,2}(1P)\gamma$
and $\chi_{b2}(2P)\gamma$ channels are also sizeable. Taking the
mass of $M_{\Upsilon_1(2D)}=10425$ MeV, we calculate
the partial decay widths. Our predicted partial widths for the
transitions $\Upsilon_1(2D)\to \chi_{b0,1,2}(1P,2P)\gamma$ compared with the
other model predictions are listed in Tab.~\ref{tab2}.
From the table, it is found that most of our predictions are compatible with
the other potential predictions in magnitude. In Ref.~\cite{Godfrey:2015dia},
the total width of $\Upsilon_1(2D)$ is predicted to be
$\Gamma_{\mathrm{tot}}\simeq 38$ keV, with this input, we estimate the
branching ratios for the dominant radiative transitions of $\Upsilon_1(2D)$, which are
\begin{eqnarray}
\mathcal{B}[\Upsilon_1(2D)\to \chi_{b0}(2P) \gamma]& \simeq &
25\%,\\
\mathcal{B}[\Upsilon_1(2D)\to \chi_{b1}(2P) \gamma]& \simeq & 18\%,\\
\mathcal{B}[\Upsilon_1(2D)\to \chi_{b0}(1P) \gamma]& \simeq &
15\%,\\
\mathcal{B}[\Upsilon_1(2D)\to \chi_{b1}(1P) \gamma]& \simeq & 7\%.
\end{eqnarray}
There may be hope for observing the missing $\Upsilon_1(2D)$ state
in the $\chi_{b0,1}(2P) \gamma$ and $\chi_{b0,1}(1P) \gamma$ channels.

\subsubsection{$\eta_{b2}(2D)$}

The main EM decay channels of $\eta_{b2}(2D)$ are $h_b(2P)\gamma$ and $h_b(1P)\gamma$. With the mass
$M_{\eta_{b2}(2D)}=10432$ MeV predicted by us,
the partial widths of the transitions
$\eta_{b2}(2D)\to h_b(1P,2P)\gamma$ are calculated. The results compared with the
other model predictions are listed in Tab.~\ref{tab2}.
It is found that the predicted partial widths roughly agree with the potential model
predictions~\cite{Godfrey:2015dia,Chao:2009,Segovia:2016xqb}.
Using the predicted total width of $\eta_{b2}(2D)$
($\Gamma_{\mathrm{tot}}\simeq 25$ keV) from~\cite{Godfrey:2015dia}, we
predict that
\begin{eqnarray}
\mathcal{B}[\eta_{b2}(2D)\to h_{b}(1P) \gamma]& \simeq &23\%,\\
\mathcal{B}[\eta_{b2}(2D)\to h_{b}(2P) \gamma]& \simeq & 62\%.
\end{eqnarray}
To determine the missing $\eta_{b2}(2D)$ state in experiments, its
transitions into the $h_{b}(1P,2P) \gamma$ channels are worth observing.

\subsection{Radiative transitions of $3P$ states}

In the past several years, obvious progress has been achieved in the
observations of the $3P$ states. In 2011, the ATLAS
Collaboration first discovered the $\chi_b(3P)$ through its
radiative transitions to $\Upsilon(1S,2S)$ with $\Upsilon(1S,2S)\to
\mu^+\mu^-$ at the LHC~\cite{Aad:2011ih}. Only a few months after
that, the $\chi_b(3P)$ state was confirmed by the D0
Collaboration~\cite{Abazov:2012gh}. Recently, the LHCb Collaboration
also carried out a precise measurement of the $\chi_b(3P)$ state,
identifying $\chi_b(3P)$ as the
$\chi_{b1}(3P)$ state~\cite{Aaij:2014hla,Aaij:2014caa}. The measured mass
of $\chi_{b1}(3P)$ is $M_{\chi_{b1}(3P)}\simeq 10516$ MeV. In our calculations,
the mass splittings are predicted to be $M_{\chi_{b2}(3P)}-M_{\chi_{b1}(3P)}\simeq 13$ MeV,
$M_{\chi_{b1}(3P)}-M_{\chi_{b0}(3P)}\simeq 25$ MeV, and
$M_{h_{b}(3P)}-M_{\chi_{b1}(3P)}\simeq 4$ MeV. Combining these
predicted mass splittings with the measured
mass of $\chi_{b1}(3P)$, we estimate the masses of $\chi_{b2}(3P)$,
$\chi_{b0}(3P)$ and $h_{b}(3P)$, which are $M_{\chi_{b2}(3P)}\simeq 10529$ MeV,
$M_{\chi_{b0}(3P)}\simeq 10491$ MeV, and $M_{h_{b}(3P)}\simeq 10520$ MeV,
respectively.

\subsubsection{$\chi_{b1}(3P)$}

The $\Upsilon(1S,2S,3S)\gamma$ are the main EM decay channels of
$\chi_{b1}(3P)$. From Tab.~\ref{tab1}, it is seen that
our predicted partial widths for these channels are
close to the recent predictions with the nonrelativistic
constituent quark model~\cite{Segovia:2016xqb}, and the predictions with the previous SNR potential
models~\cite{Chao:2009}. Furthermore, taking the total width of
$\chi_{b1}(3P)$, $\Gamma_{\mathrm{tot}}\simeq 117$ keV,
predicted in Ref.~\cite{Godfrey:2015dia} as an input, we
estimate that
\begin{eqnarray}
\mathcal{B}[\chi_{b1}(3P)\to \Upsilon(1S)\gamma]& \simeq & 5.4\%,\\
\mathcal{B}[\chi_{b1}(3P)\to \Upsilon(2S)\gamma]& \simeq &4.8\%,\\
\mathcal{B}[\chi_{b1}(3P)\to \Upsilon(3S)\gamma]& \simeq &8.8\%.
\end{eqnarray}
These large branching ratios may explain why $\chi_{b}(3P)$ is
discovered through its radiative transitions into $\Upsilon(1S,2S)$.

Taking the masses of $2D$ waves calculated by us, we predict the partial widths
for the transitions $\chi_{b1}(3P)\to \Upsilon_{1,2}(2D)\gamma$.
Our results are listed in Tab.~\ref{tab1}. From the table, it is found that
our results are close to the potential model
predictions~\cite{Godfrey:2015dia,Chao:2009}. Similarly, with the
predicted total width $\chi_{b1}(3P)$ from~\cite{Godfrey:2015dia}, we
estimate that
\begin{eqnarray}
\mathcal{B}[\chi_{b1}(3P)\to \Upsilon_1(2D)\gamma]& \simeq & 9.0\times 10^{-3},\\
\mathcal{B}[\chi_{b1}(3P)\to \Upsilon_2(2D)\gamma]& \simeq & 8.0\times 10^{-3}.
\end{eqnarray}
The sizeable branching ratios of $\mathcal{B}[\chi_{b1}(3P)\to \Upsilon_{1,2}(2D)\gamma]$
indicate that one may discover the missing $D$-wave states
$\Upsilon_1(2D)$ and $\Upsilon_2(2D)$ through the radiative
transition chains $\chi_{b1}(3P)\to \Upsilon_{1,2}(2D)\gamma\to \chi_{b1}(1P,2P)\gamma\gamma \to \Upsilon(1S,2S)\gamma\gamma\gamma$.
We further estimate the branching ratios for these decay chains. The results are listed
in Tab.~\ref{3p1chain}. It is found that the important chains involving $\Upsilon_1(2D)$ are
$\chi_{b1}(3P)\to \Upsilon_{1}(2D)\gamma\to \chi_{b1}(2P,1P)\gamma\gamma \to \Upsilon(1S,2S)\gamma\gamma\gamma$
[$\mathcal{B}\sim \mathcal{O} (10^{-4})$]. While the important chains involving $\Upsilon_2(2D)$ are
$\chi_{b1}(3P)\to \Upsilon_{2}(2D)\gamma\to \chi_{b1}(2P)\gamma\gamma \to \Upsilon(2S)\gamma\gamma\gamma$
[$\mathcal{B}\simeq 4.6\times 10^{-4}$], $\chi_{b1}(3P)\to \Upsilon_{2}(2D)\gamma\to \chi_{b1}(1P)\gamma\gamma \to \Upsilon(1S)\gamma\gamma\gamma$
[$\mathcal{B}\simeq 4.6\times 10^{-4}$], and $\chi_{b1}(3P)\to \Upsilon_{2}(2D)\gamma\to \chi_{b1}(2P)\gamma\gamma \to \Upsilon(1S)\gamma\gamma\gamma$ [$\mathcal{B}\simeq 3.2\times 10^{-4}$].

\subsubsection{$\chi_{b2}(3P)$}

With $M_{\chi_{b2}(3P)}=10529$ MeV for the
$\chi_{b2}(3P)$ state, we calculate its radiative decay properties.
Our results are listed in Tab.~\ref{tab1}.
For comparison, the other model predictions are also listed in the same table.
It is found that the radiative decay
ratios of $\chi_{b2}(3P)$ into the $1D$-wave states are negligibly
small, while the partial widths for the transitions
$\chi_{b2}(3P)\to \Upsilon(1S,2S,3S)\gamma$ and $\chi_{b2}(3P)\to
\Upsilon_3(2D)\gamma$ are sizeable. Most of our results are consistent
with the other predictions. Taking the total width of $\chi_{b2}(3P)$,
$\Gamma_{\mathrm{tot}}\simeq 247$ keV, predicted in Ref.~\cite{Godfrey:2015dia}
as an input, we estimate that
\begin{eqnarray}
\mathcal{B}[\chi_{b2}(3P)\to \Upsilon(1S)\gamma]& \simeq &3.3\%,\\
\mathcal{B}[\chi_{b2}(3P)\to \Upsilon(2S)\gamma]& \simeq &2.7\%,\\
\mathcal{B}[\chi_{b2}(3P)\to \Upsilon(3S)\gamma]& \simeq &4.4\%.
\end{eqnarray}
These fairly large branching ratios indicate the missing $\chi_{b2}(3P)$
state is most likely to be established via the radiative
decays $\chi_{b2}(3P)\to \Upsilon(1S,2S,3S) \gamma$.
Furthermore, we find that the branching ratio
\begin{eqnarray}
\mathcal{B}[\chi_{b2}(3P)\to \Upsilon_3(2D)\gamma]& \simeq & 1.9\%
\end{eqnarray}
is sizeable. Thus, $\chi_{b2}(3P)$ might be a good source when looking for the missing
$\Upsilon_3(2D)$. According to our analysis, the important radiative decay chains involving
$\Upsilon_3(2D)$ are $\chi_{b2}(3P)\to \Upsilon_3(2D)\gamma\to \chi_{b2}(2P)\gamma\gamma
\to\Upsilon(1S,2S)\gamma\gamma\gamma$, and their combined branching ratios can reach up to
$\mathcal{B}\simeq 1.3\times 10^{-3}$.

\subsubsection{$\chi_{b0}(3P)$}

With the predicted mass $M_{\chi_{b0}(3P)}=10491$ MeV for the
$\chi_{b0}(3P)$ state, we calculate its radiative decay properties.
Our results are listed in Tab.~\ref{tab1}. It is found that the partial
radiative decay widths of $\chi_{b0}(3P)$ into the $S$-wave states
$\Upsilon(1S,2S,3S)$ are comparable to those of $\chi_{b1,2}(3P)$.
In Ref.~\cite{Godfrey:2015dia}, the total width of $\chi_{b0}(3P)$
is predicted to be $\Gamma_{\mathrm{tot}}\simeq 2.5$ MeV,
with which we estimate that
\begin{eqnarray}
\mathcal{B}[\chi_{b0}(3P)\to \Upsilon(1S)\gamma]& \simeq &7.5\times 10^{-4},\\
\mathcal{B}[\chi_{b0}(3P)\to \Upsilon(2S)\gamma]& \simeq &1.0\times 10^{-4},\\
\mathcal{B}[\chi_{b0}(3P)\to \Upsilon(3S)\gamma]& \simeq &3.2\times 10^{-4}.
\end{eqnarray}
These branching ratios are about an order of magnitude smaller than those of
$\mathcal{B}[\chi_{b1,2}(3P)\to \Upsilon(1S,2S,3S)\gamma]$, which may
indicate that $\chi_{b0}(3P)$ is relatively difficult to observe in the
$\Upsilon(1S,2S,3S)\gamma$ channels.

\subsubsection{$h_{b}(3P)$}

For the singlet $h_{b}(3P)$ state, with the predicted mass
$M_{h_{b}(3P)}=10520$ MeV, we calculate the radiative decay
properties. Our results are listed in Tab.~\ref{tab1}. The EM decays of
$h_{b}(3P)$ are dominated by the $\eta_b(3S)\gamma $ channel, while
the partial widths into the $\eta_b(1S,2S)\gamma $ and
$\eta_{b2}(2D)\gamma $ channels are sizeable as well. Our predicted
partial decay widths into the $S$-wave states are the same order
of those from various potential models
~\cite{Godfrey:2015dia,Chao:2009,Segovia:2016xqb}
(see Tab.~\ref{tab1}). Taking the predicted
width of $h_{b}(3P)$, $\Gamma_{\mathrm{tot}}\simeq 83$ keV,
from Ref.~\cite{Godfrey:2015dia} as an input, we obtain
\begin{eqnarray}
\mathcal{B}[h_{b}(3P)\to \eta_b(1S)\gamma]& \simeq &12.9\%,\\
\mathcal{B}[h_{b}(3P)\to \eta_b(2S)\gamma]& \simeq &9.2\%,\\
\mathcal{B}[h_{b}(3P)\to \eta_b(3S)\gamma]& \simeq &17.0\%.
\end{eqnarray}
To look for the missing $h_{b}(3P)$ state, the
transitions $h_{b}(3P)\to \eta_b(1S,2S)\gamma$ are worth observing.

\subsection{Radiative transitions of $4S$ states}

$\Upsilon(4S)$ is established in experiments. Its mass and
width are $M_{\Upsilon(4S)}\simeq 10579$ MeV and $\Gamma\simeq20.5$ MeV, respectively.
However, the $\eta_b(4S)$ is still missing.
We predict their radiative properties. The results compared with
the other predictions are listed in Tab.~\ref{tab1}.
From the table, it is found that obvious model dependencies exist in these predictions.
Our calculations give relatively large decay rates for the $\Upsilon(4S)\to\chi_{bJ}(3P)\gamma$ transitions.
Thus, the missing $\chi_{bJ}(3P)$ states might be produced by the radiative decay chains of
$\Upsilon(4S)\to \chi_{bJ}(3P)\gamma\to \Upsilon(1S,2S,3S)\gamma\gamma$.
Combining the predicted branching ratios of $\chi_{bJ}(3P)\to \Upsilon(1S,2S,3S) \gamma$
and $\Upsilon(4S)\to \chi_{bJ}(3P)\gamma$,
we further estimate the combined branching ratios, which have been listed in
Tab.~\ref{EM4s}. From the table, one can see that the most prominent
two-photon decay chains are
$\Upsilon(4S)\to \chi_{b1}(3P)\gamma\to \Upsilon(1S,2S,3S)\gamma\gamma$
[$\mathcal{B}\sim \mathcal{O}(10^{-5})$], followed by
$\Upsilon(4S)\to \chi_{b2}(3P)\gamma\to \Upsilon(1S,2S,3S)\gamma\gamma$
[$\mathcal{B}\sim \mathcal{O}(10^{-6})$]. There are
few chances for $\chi_{b0}(3P)$ to be observed in the radiative decay
chains of $\Upsilon(4S)$.

\section{Summary}\label{sum}

In the nonrelativistic screened potential quark model framework, we study
the bottomonium spectrum. The radial Schr\"{o}dinger
equation is solved with the three-point difference
central method, where the spin-dependent potentials are dealt with non-perturbatively.
In our calculations, the corrections of the spin-dependent interactions to
the wave functions are successfully included as well.
It is found that the corrections of spin-dependent interactions to the wave functions of
the $S$-wave and $^3P_{0,1}$ states are notably big.
The bottomonium spectrum predicted within our approach is in a global agreement with the experimental
data.

Moreover, using the obtained wave functions we study
the EM transitions of $nS$ ($n\leq 4$), $nP$ ($n\leq 3$), and $nD$ ($n\leq 2$)
bottomonium states with a nonrelativistic EM
transition operator widely applied to meson photoproduction reactions, in which the effects
of binding potential between quarks are considered, and the possible higher EM multipole
contributions are included. It is found that (i) except for some M1 transitions,
our predictions for the EM transitions are in good agreement with the experimental data.
(ii) The corrections of the spin-dependent interactions are important to
understand some EM transitions. For example, the EM transitions
of $\Upsilon(3S)\to \chi_{b1,2}(1P)\gamma$, which were not well
understood in previous studies, can be reasonably explained in the present work by considering
the corrections of the spin-dependent interactions to the wave functions.
(iii) Strong model dependencies exist in various model predictions of some transition widths,
especially for the partial width ratios. To test the various model predictions more
observations are expected to be carried out in forthcoming experiments.

Additionally, we discuss the observations of the missing bottomonium states by using radiative
transitions. (i) We suggest our experimental colleagues observe the three-photon decay chains
$\chi_{b2}(2P)\to \Upsilon_3(1D) \gamma \to \chi_{b2}(1P)\gamma\gamma
\to \Upsilon(1S)\gamma\gamma\gamma$ [$\mathcal{B}\sim\mathcal{O}(10^{-3})$] and $h_{b}(2P)\to\eta_{b2}(1D)  \gamma \to h_b(1P)\gamma\gamma\to \eta_b\gamma\gamma\gamma$ ($\mathcal{B}\simeq 1.4\%$), where the missing $\Upsilon_3(1D)$ and $\eta_{b2}(1D)$
states are most likely to be observed.
(ii) We also suggest our experimental colleagues observe the following three-photon decay chains:
$\chi_{b1}(3P)\to \Upsilon_{1}(2D)\gamma\to \chi_{b1}(2P,1P)\gamma\gamma \to \Upsilon(1S,2S)\gamma\gamma\gamma$
[$\mathcal{B}\sim \mathcal{O} (10^{-4})$], $\chi_{b1}(3P)\to \Upsilon_{2}(2D)\gamma\to \chi_{b1}(2P)\gamma\gamma \to \Upsilon(2S)\gamma\gamma\gamma$
[$\mathcal{B}\simeq 4.6\times 10^{-4}$], $\chi_{b1}(3P)\to \Upsilon_{2}(2D)\gamma\to \chi_{b1}(1P)\gamma\gamma \to \Upsilon(1S)\gamma\gamma\gamma$
[$\mathcal{B}\simeq 4.6\times 10^{-4}$], and $\chi_{b1}(3P)\to \Upsilon_{2}(2D)\gamma\to \chi_{b1}(2P)\gamma\gamma \to \Upsilon(1S)\gamma\gamma\gamma$ [$\mathcal{B}\simeq 3.2\times 10^{-4}$], where the missing $\Upsilon_{1}(2D)$ and $\Upsilon_{2}(2D)$ states might have chances to be observed. (iii) The missing $\chi_{bJ}(3P)$ states might be produced via the radiative transitions of $\Upsilon(4S)$.
The most prominent decay chains are $\Upsilon(4S)\to \chi_{b1}(3P)\gamma\to \Upsilon(1S,2S,3S)\gamma\gamma$
[$\mathcal{B}\sim \mathcal{O}(10^{-5})$], followed by
$\Upsilon(4S)\to \chi_{b2}(3P)\gamma\to \Upsilon(1S,2S,3S)\gamma\gamma$
[$\mathcal{B}\sim \mathcal{O}(10^{-6})$].

The LHC and Belle experiments have demonstrated the ability to observe
and measure the properties of bottomonium mesons.
In the near future, more missing bottomonium states are to be discovered
and more decay channels will be measured in experiments. We expect that
our theoretical predictions in this paper will be helpful for experimental
exploration of the bottomonium mesons.

\begin{table*}[htb]
\caption{ Partial widths of the radiative transitions
for the $nS$- and $nP$-wave ($n=2,3$) bottomonium states.
For comparison, the
measured values from the PDG~\cite{PDG}, the predictions from the relativistic quark
model~\cite{Ebert:2003}, relativized quark model (GI model)~\cite{Godfrey:2015dia}, nonrelativistic
constituent quark model (NR model)~\cite{Segovia:2016xqb}, and the previous screened potential
model (SNR model)~\cite{Chao:2009} are listed in the table as well. SNR$_{0}$ and SNR$_{1}$ stand for
the results calculated by the zeroth-order wave functions and the first-order relativistically
corrected wave functions with the screened potential model~\cite{Chao:2009}, respectively.
}\label{tab1}
\begin{tabular}{c|c|cccc|ccccc|cc|cc}  \hline\hline
 Initial       & Final & \multicolumn{4}{|c|} {\underline{$E_{\gamma}$  (MeV)}  } & \multicolumn{5}{|c|} {\underline{$\Gamma_{\mathrm{E1}}$  (keV)}} & \multicolumn{2}{|c|} {\underline{$\Gamma_{\mathrm{EM}}$  (keV)}}  \\
   state             & state            & Ref.\cite{Ebert:2003}& SNR$_{0,1}$\cite{Chao:2009}& GI~\cite{Godfrey:2015dia} &  ours &   Ref.\cite{Ebert:2003}& SNR$_{0}$\cite{Chao:2009}& SNR$_{1}$\cite{Chao:2009}& GI~\cite{Godfrey:2015dia} & NR~\cite{Segovia:2016xqb}  & Ours &PDG~\cite{PDG} \\
\hline
$\Upsilon(2 ^3S_{1})$~~~~~   &$\chi_{b2}(1 ^3P_{2})$       & 109 & 110 &110&  110 & 2.46 & 2.62 & 2.46 &1.88 & 2.08   & 2.62  &$2.29\pm0.20$\\
                             &$\chi_{b1}(1 ^3P_{1})$       & 130 & 130 &129&  129 & 2.45 & 2.54 & 2.08 &1.63 & 1.84   & 2.17  &$2.21\pm0.19$\\
                             &$\chi_{b0}(1 ^3P_{0})$       & 162 & 163 &163&  163 & 1.62 & 1.67 & 1.11 &0.91 & 1.09   & 1.09  &$1.22\pm0.11$\\
$\eta_{b}(2 ^1S_{0})$ &$h_{b}(1 ^1P_{1})$           & 98~ &  83 &99&   99 & 3.09 & 6.10 & 5.57 &2.48 & 2.85   & 3.41  & \\
\hline
$\Upsilon(3 ^3S_{1})$        &$\chi_{b2}(2 ^3P_{2})$       & 86~ &  86 &86&   86 & 2.67 & 3.23 & 3.04 &2.30 & 2.56   & 3.16  &$2.66\pm0.27$\\
                             &$\chi_{b1}(2 ^3P_{1})$       & 100 &  99 &100&  100 & 2.41 & 2.96 & 2.44 &1.91 & 2.13   & 2.61  &$2.56\pm0.26$\\
                             &$\chi_{b0}(2 ^3P_{0})$       & 123 & 122 &121&  121 & 1.49 & 1.83 & 1.23 &1.03 & 1.21   & 1.21  &$1.20\pm0.12$\\
                             &$\chi_{b2}(1 ^3P_{2})$       & 433 & 434 &434&  434 & 0.097& 0.25 & 1.26 &0.45 & 0.083  & 0.14 &$0.20\pm0.03$\\
                             &$\chi_{b1}(1 ^3P_{1})$       & 453 & 452 &452&  452 & 0.067& 0.17 & 0.14 &0.05 & 0.16  & 0.0005 &$0.018\pm0.010$\\
                             &$\chi_{b0}(1 ^3P_{0})$       & 484 & 484 &484&  484 & 0.027& 0.07 & 0.05 &0.01 & 0.15  & 0.097 &$0.055\pm0.010$\\
$\eta_{b}(3 ^1S_{0})$        &$h_{b}(2 ^1P_{1})$           & 73  &  74 &77&   78 & 2.78 & 11.0 & 10.1 &2.96 & 2.60   & 4.25  &  \\
                             &$h_{b}(1 ^1P_{1})$           & 427 & 418 &429&  429 & 0.348& 1.24 & 5.68 &1.30 & 0.0084  & 0.67 &  \\
\hline
$\chi_{b2}(2 ^3P_{2})$~~~~~  &$\Upsilon(1 ^3D_{3})$        & 108 & 113 &97&  101 & 2.35 & 3.33 & 3.13 &1.5 & 2.06   & 2.51  & \\
                             &$\Upsilon(1 ^3D_{2})$        & 111 & 117 &104&  104 & 0.449& 0.66 & 0.58 &0.3 & 0.35  & 0.42 & \\
                             &$\Upsilon(1 ^3D_{1})$        & 117 & 123 &113&  111 & 0.035& 0.05 & 0.04 &0.03& 0.021 & 0.026& \\
                             &$\Upsilon(2 ^3S_{1})$        & 243 & 243 &243&  243 & 16.7 & 18.8 & 14.2 &14.3& 17.50   & 15.3  &$15.1\pm5.6$ \\
                             &$\Upsilon(1 ^3S_{1})$        & 776 & 777 &777&  777 & 8.02 & 13.0 & 12.5 &8.4 & 11.38   & 12.5  &$9.8\pm2.3$ \\
$\chi_{b1}(2 ^3P_{1})$~~~~~  &$\Upsilon(1 ^3D_{2})$        & 98~ & 104 &91&   91 & 1.56 & 2.31 & 2.26 &1.2 & 1.26   & 0.50  & \\
                             &$\Upsilon(1 ^3D_{1})$        & 104 & 110 &100&  98 & 0.615& 0.92 & 0.84 &0.5 & 0.41  & 0.56 & \\
                             &$\Upsilon(2 ^3S_{1})$        & 230 & 230 &229&  229 & 14.7 & 15.9 & 13.8 &13.3& 15.89   & 15.3  & $19.4\pm5.0$ \\
                             &$\Upsilon(1 ^3S_{1})$        & 764 & 764 &764&  764 & 7.49 & 12.4 & 8.56 &5.5 & 9.13   & 10.8  & $8.9\pm2.2$\\
$\chi_{b0}(2 ^3P_{0})$~~~~~  &$\Upsilon(1 ^3D_{1})$        & 81~ & ~87 &78&   78 & 1.17 & 1.83 & 1.85 &1.0 & 0.74   & 1.77  & \\
                             &$\Upsilon(2 ^3S_{1})$        & 207 & 207 &208&  208 & 11.0 & 11.7 & 11.6 &10.9& 12.80   & 14.4  & \\
                             &$\Upsilon(1 ^3S_{1})$        & 743 & 743 &744&  744 & 6.79 & 11.4 & 4.50 &2.5 & 5.44   & 5.54  & \\
$h_{b}(2 ^1P_{1})$~~~~~      &$\eta_{b2}(1 ^1D_{2})$       & 102 & 104 &95&  95 & 2.43 & 7.74 & 7.42 &1.7 & 5.36  & 2.24  & \\
                             &$\eta_{b}^\prime(2 ^1S_{0})$ & 262 & 266 &258&  258 & 21.4 & 24.7 & 15.3 &14.1& 17.60  & 16.2  & \\
                             &$\eta_{b}(1 ^1S_{0})$        & 820 & 831 &826&  826 & 9.36 & 15.9 & 18.0 &13.0& 14.90   & 16.1  & \\
\hline
$\chi_{b2}(3 ^3P_{2})$~~~~~  &$\Upsilon(2 ^3D_{3})$        &     &  97 &73&   93 &      & 5.05 & 4.69 &1.5 & 4.16   & 4.60  & \\
                             &$\Upsilon(2 ^3D_{2})$        &     & 101 &79&  97 &       & 1.02 & 0.89 &0.32& 0.79  & 0.78 & \\
                             &$\Upsilon(2 ^3D_{1})$        &     & 107 &87&  103 &       & 0.08 & 0.07 &0.027& 0.18 & 0.049& \\
                             &$\Upsilon(1 ^3D_{3})$        &     & 377 &350&  365 &      &    0 & 0.05 &0.046& 0.21  & 0.12 & \\
                             &$\Upsilon(1 ^3D_{2})$        &     &   & &  358 &      &      &      &     & 0.044 & 0.068 & \\
                             &$\Upsilon(1 ^3D_{1})$        &     &   & &  366 &      &      &      &     & 0.0034 & 0.047& \\
                             &$\Upsilon(3 ^3S_{1})$        &     & 183 &172&  173 &      & 15.6 & 11.1 &9.3& 10.38   & 10.8  & \\
                             &$\Upsilon(2 ^3S_{1})$        &     & 504 &493&  494 &      & 6.00 & 6.89 &4.5& 5.62   &6.72 & \\
                             &$\Upsilon(1 ^3S_{1})$        &     & 1024&1014&  1014&      & 7.09 & 6.76 &2.8& 5.65   & 8.17  & \\
$\chi_{b1}(3 ^3P_{1})$~~~~~  &$\Upsilon(2 ^3D_{2})$        &     & 86  &67&    84&      & 3.10 & 2.98 &1.1& 3.34   & 0.94 & \\
                             &$\Upsilon(2 ^3D_{1})$        &     & 92  &75&    90&      & 1.26 & 1.13 &0.47& 1.26  & 1.07 & \\
                             &$\Upsilon(1 ^3D_{2})$        &     & 366 &346&  346 &      &    0 & 0.09 &0.08& 0.11  & 0.015 & \\
                             &$\Upsilon(1 ^3D_{1})$        &     & 372 &355&  355 &      &    0 & 0.00 &0.007& 0.048 & 0.010& \\
                             &$\Upsilon(3 ^3S_{1})$        &     & 167 &160&  160 &      & 12.0 & 9.97 &8.4 & 9.62   & 10.3  & \\
                             &$\Upsilon(2 ^3S_{1})$        &     & 489 &481&  481 &      & 5.48 & 5.39 &3.1 & 4.58   & 5.63  & \\
                             &$\Upsilon(1 ^3S_{1})$        &     & 1010&1003&  1003&      & 6.80 & 3.39 &1.3 & 4.17   &6.41  & \\
$\chi_{b0}(3 ^3P_{0})$~~~~~  &$\Upsilon(2 ^3D_{1})$        &     &     &59&  68  &      &      &      &1.0 & 3.50   & 2.20  & \\
                             &$\Upsilon(1 ^3D_{1})$        &     & 351 &339&  341 &      &   0  & 0.17 &0.20& 0.036  & 0.15 & \\
                             &$\Upsilon(3 ^3S_{1})$        &     & 146 &144&  135 &      &7.88  & 7.67 &6.9 & 8.50   & 7.95  & \\
                             &$\Upsilon(2 ^3S_{1})$        &     & 468 &466&  458 &      &4.80  & 3.67 &1.7 & 2.99   & 2.55  & \\
                             &$\Upsilon(1 ^3S_{1})$        &     & 990 &988&  980 &      &6.41  & 0.86 &0.3 & 1.99   & 1.87 & \\
$h_{b}(3 ^1P_{1})$~~~~~      &$\eta_{b2}(2 ^1D_{2})$       &     &     &69&   88 &      &      &      &1.6 &  4.72  & 4.21  & \\
                             &$\eta_{b2}(1 ^1D_{2})$       &     & 370 &348&  360 &      &   0  & 0.24 &0.081& 0.35  & 0.17 & \\
                             &$\eta_{b}(3 ^1S_{0})$        &     & 196 &180&  194 &      &19.2  & 11.6 &8.9 & 12.27   & 14.1  & \\
                             &$\eta_{b}(2 ^1S_{0})$        &     & 528 &507&  508 &      &6.89  & 10.3 &8.2 & 6.86   & 7.63  & \\
                             &$\eta_{b}(1 ^1S_{0})$        &     & 1078&1061&  1062&      &8.27  & 9.46 &3.6 & 7.96   & 10.7  & \\
 \hline\hline

 \end{tabular}
 \end{table*}

\begin{table*}[htb]
\caption{ Partial widths of the radiative transitions
for the $1P$-, $1D$- and $2D$-wave bottomonium states. For comparison, the
predictions from the relativistic quark model~\cite{Ebert:2003}, relativized quark model (GI model)~\cite{Godfrey:2015dia}, nonrelativistic
constituent quark model (NR model)~\cite{Segovia:2016xqb}, and previous screened potential model (SNR model)~\cite{Chao:2009} are
listed in the table as well. SNR$_{0}$ and SNR$_{1}$ stand for
the results calculated by the zeroth-order wave functions and the first-order relativistically
corrected wave functions with the screened potential model~\cite{Chao:2009}, respectively.}\label{tab2}
\begin{tabular}{c|c|cccc|ccccc|cccc}  \hline\hline
 Initial meson       & Final meson & \multicolumn{4}{|c|} {\underline{$E_{\gamma}$  (MeV)}  } & \multicolumn{5}{|c|} {\underline{$\Gamma_{\mathrm{E1}}$  (keV)}} & \multicolumn{1}{|c} {\underline{$\Gamma_{\mathrm{EM}}$  (keV)}}  \\
   state             & state            & Ref.\cite{Ebert:2003}& SNR$_{0,1}$\cite{Chao:2009}& GI~\cite{Godfrey:2015dia} &  ours &   Ref.\cite{Ebert:2003}& SNR$_{0}$\cite{Chao:2009}& SNR$_{1}$\cite{Chao:2009}& GI~\cite{Godfrey:2015dia} & NR~\cite{Segovia:2016xqb} & Ours \\
\hline
\hline
$\chi_{b2}(1 ^3P_{2})$ &$\Upsilon(1 ^3S_{1})$        & 442 & 442 &442&  442 & 40.2 & 38.2 & 32.6 &32.8 & 39.15    & 31.8   &  \\
$\chi_{b1}(1 ^3P_{1})$  &                             & 422 & 423 &424&  424 & 36.6 & 33.6 & 30.0 &29.5 & 35.66    & 31.9   &  \\
$\chi_{b0}(1 ^3P_{0})$  &                             & 391 & 391 &391&  391 & 29.9 & 26.6 & 24.3 &23.8 & 28.07  & 27.5  &  \\
$h_{b}(1 ^1P_{1})$     &$\eta_{b}(1 ^1S_{0})$        & 480 & 501 &488&  488 & 52.6 & 55.8 & 36.3 &35.7 & 43.66    & 35.8   &  \\
\hline
$\Upsilon(1 ^3D_{3})$      &$\chi_{b2}(1 ^3P_{2})$        & 244  & 240 &257& 253  & 24.6 & 26.4 & 24.5 &24.3 &  24.74 &$32.1$  &  \\
                           &$\chi_{b1}(1 ^3P_{1})$      &      &    &  & 271  &      & ~~~~ & ~~~~ &     & $0$   &$1.1\times10^{-2}$ &  \\
                           &$\chi_{b0}(1 ^3P_{0})$      &      &    &  & 304  &      & ~~~~ & ~~~~ &     & $0$   &$9.2\times10^{-5}$ &  \\
$\Upsilon(1 ^3D_{2})$      &$\chi_{b2}(1 ^3P_{2})$      & 241  & 236&249 & 249  & 6.35 & 6.29 & 5.87 &5.6  &   6.23              &$7.23$             &  \\
                           &$\chi_{b1}(1 ^3P_{1})$      & 262  & 255&267 & 267  & 23.3 & 23.8 & 19.8 &19.2 &   21.95              &$21.8$             &  \\
                           &$\chi_{b0}(1 ^3P_{0})$      &      &    &  & 300  &      & ~~~~ & ~~~~ &     & $0$   &$0.83\times10^{-2}$         &  \\
$\Upsilon(1 ^3D_{1})$      &$\chi_{b2}(1 ^3P_{2})$      & 235  & 230&240 & 242  & 0.69 & 0.65 & 0.61 &0.56 &   0.65              &$1.02$             &  \\
                           &$\chi_{b1}(1 ^3P_{1})$      & 256  & 249&259 & 261  & 12.7 & 12.3 & 10.3 &9.7  &   12.29              &$13.3$             &  \\
                           &$\chi_{b0}(1 ^3P_{0})$      & 280  & 282&292 & 294  & 23.4 & 23.6 & 16.7 &16.5 &   20.98               &$19.8$             &  \\
$\eta_{b2}(1 ^1D_{2})$     &$h_{b}(1 ^1P_{1})$          & 254  & 246&263 & 262  & 28.4 & 42.3 & 36.5 &24.9 &   17.23               &$30.3$             &  \\
\hline
$\Upsilon(2 ^3D_{3})$      &$\chi_{b2}(1 ^3P_{2})$      &      & 517&529 & 511  &      & 4.01 & 3.73 &2.6  &  3.80                &$5.22$             &  \\
                           &$\chi_{b1}(1 ^3P_{1})$      &      &    & & 535  &      & ~~~~ & ~~~~ &     &  0               &$0.16$            &  \\
                           &$\chi_{b0}(1 ^3P_{0})$      &      &    &  & 567  &      & ~~~~ & ~~~~ &     &  0               &$0.08$            &  \\
                           &$\chi_{b2}(2 ^3P_{2})$      &      & 172&184 & 166  &      & 18.0 & 15.9 &16.4 &  10.70                &$17.0$             &  \\
                           &$\chi_{b1}(2 ^3P_{1})$      &      &    &  & 185  &      & ~~~~ & ~~~~ &     & $0$   &$0.34\times10^{-2}$           &  \\
                           &$\chi_{b0}(2 ^3P_{0})$      &      &    &  & 207  &      & ~~~~ & ~~~~ &     & $0$   &$0.66\times10^{-3}$           &  \\
$\Upsilon(2 ^3D_{2})$      &$\chi_{b2}(1 ^3P_{2})$      &      & 513&523 & 507  &      & 0.98 & 0.68 &0.4  &   0.80              &$1.11$             &  \\
                           &$\chi_{b1}(1 ^3P_{1})$      &      & 531&541 & 525  &      & 3.26 & 4.46 &2.6  &   3.43               &$4.00$             &  \\
                           &$\chi_{b0}(1 ^3P_{0})$      &      & ~~~&  & 555  &      & ~~~~ & ~~~~ & & $0$   &$0.89\times10^{-2}$         &  \\
                           &$\chi_{b2}(2 ^3P_{2})$      &      & 168&178 & 162  &      & 4.17 & 3.82 &3.8  &   2.55               &$3.75$             &  \\
                           &$\chi_{b1}(2 ^3P_{1})$      &      & 181&192 & 175  &      & 15.7 & 12.1 &12.7 &   9.10               &$11.4$             &  \\
                           &$\chi_{b0}(2 ^3P_{0})$      &      & ~~~&  & 197  &      & ~~~~ & ~~~~ &     & $0$    &$1.7\times10^{-3}$         &  \\
$\Upsilon(2 ^3D_{1})$      &$\chi_{b2}(1 ^3P_{2})$      &      & 507&516 & 500  &      & 0.11 & 0.05 &0.9 &  0.061                &$0.44$             &  \\
                           &$\chi_{b1}(1 ^3P_{1})$      &      & 525&534 & 518  &      & 1.76 & 1.87 &2.9 &   1.58             &$2.17$            &  \\
                           &$\chi_{b0}(1 ^3P_{0})$      &      & 557&566 & 551  &      & 2.79 & 6.20 &1.6 &   3.52               &$5.56$             &  \\
                           &$\chi_{b2}(2 ^3P_{2})$      &      & 162&171 & 155  &      & 0.42 & 0.39 &0.4  &  0.24               &$0.47$            &  \\
                           &$\chi_{b1}(2 ^3P_{1})$      &      & 175&184 & 167  &      & 7.87 & 6.35 &6.5  &  4.84                &$6.74$             &  \\
                           &$\chi_{b0}(2 ^3P_{0})$      &      & 198&206 & 190  &      & 15.1 & 9.49 &10.6 &  8.35                &$9.58$             &  \\
$\eta_{b2}(2 ^1D_{2})$     &$h_{b}(1 ^1P_{1})$          &      & 522&536 & 519  &      & 6.19 & 7.30 &3.0 &   4.15               &$5.66$             &  \\
                           &$h_{b}(2 ^1P_{1})$          &      & 181&188 & 171  &      & 31.3 & 25.4 &16.5 &   11.66               &$15.6$~            &  \\
\hline\hline
\end{tabular}
\end{table*}

\begin{table*}[htb]
\caption{ Partial widths of the radiative transitions for the higher $4S$ states. For comparison, the predictions from the relativized quark model (GI model)~\cite{Godfrey:2015dia}, nonrelativistic
constituent quark model (NR model)~\cite{Segovia:2016xqb}, and the previous screened potential
model (SNR model)~\cite{Chao:2009} are listed in the table as well. SNR$_{0}$ and SNR$_{1}$ stand for
the results calculated by the zeroth-order wave functions and the first-order relativistically
corrected wave functions with the screened potential model~\cite{Chao:2009}, respectively. }\label{EM4s}
\begin{tabular}{c|c|ccc|ccc|cccccccc}  \hline\hline
 Initial        & Final  & \multicolumn{3}{|c|} {\underline{$E_{\gamma}$ (MeV)}} & \multicolumn{3}{|c|} {\underline{$\Gamma_{\mathrm{E1}}$ (keV)}}
 & \multicolumn{1}{|c} {\underline{$\Gamma_{\mathrm{EM}}$ (keV)}}  \\
   state             & state            & SNR~\cite{Chao:2009}&GI~\cite{Godfrey:2015dia}&    Ours& SNR$_{0}$/SNR$_{1}$~\cite{Chao:2009} &  NR~\cite{Segovia:2016xqb} &GI~\cite{Godfrey:2015dia}&   Ours  \\
\hline
$\Upsilon(4S)$  &$\chi_{b2}(1P)$     & 646 &       &  646      &0.14/0.56 &0.012  &      &0.66  \\
                &$\chi_{b1}(1P)$     & 664 &       &  664      &0.10/0.20 &0.047  &      &0.017   \\
                &$\chi_{b0}(1P)$     & 695 &       &  695      &0.04/0.001 &0.059  &      &0.14    \\
                &$\chi_{b2}(2P)$     & 306 &       &  305      &0.14/0.56  &0.11   &      &0.34   \\
                &$\chi_{b1}(2P)$     & 319 &       &  319      &0.09/0.001 &0.18   &      &0.024   \\
                &$\chi_{b0}(2P)$     & 341 &       &  340      &0.04/0.21  &0.17   &      &0.44    \\
                &$\chi_{b2}(3P)$     & 40  &51     &  50       &0.55/0.52  &1.45   &0.82  &4.4   \\
                &$\chi_{b1}(3P)$     & 55  &63     &  64       &0.91/0.74  &1.17   &0.84  &4.9   \\
                &$\chi_{b0}(3P)$     & 77  &79     &  89       &0.82/0.54  &0.61   &0.48  &3.4   \\
\hline
$\eta_b(4S)$       &$h_{c}(1P)$     & 669 &       &  663      &0.90/5.64   &       &      &1.98  \\
                   &$h_{c}(2P)$     & 334 &       &  319      &0.95/2.16   &       &      &1.56  \\
                   &$h_{c}(3P)$     & 67  & 48    &  65       &1.24/5.68   &       &1.24  &17.4  \\
\hline

\end{tabular}
\end{table*}

\begin{table*}[htb]
\begin{center}
\caption{ Three-photon decay chains of $3^3P_2$. The combined branching fractions of the chain are defined by $\mathcal{B}=\mathcal{B}_1\times \mathcal{B}_2\times \mathcal{B}_3$ with $\mathcal{B}_1=\mathcal{B}[3^3P_1\to 2^3D_J \gamma]$, $\mathcal{B}_2=\mathcal{B}[2^3D_J \to m^3P_J\gamma]$, and $\mathcal{B}_3=\mathcal{B}[m^3P_J\to \Upsilon(1S,2S) \gamma]$.  }\label{3p1chain}
\begin{tabular}{c|c|c|c|ccccc}
\hline\hline
  Decay chain    & $\mathcal{B}_1$ & $\mathcal{B}_2$&$\mathcal{B}_3$ &$\mathcal{B}$\\
 \hline
    $ 3^3P_1\to 2^3D_1\to 2^3P_0\to \Upsilon(2S)$   &$9.0\times 10^{-3}$ &$25\%$  &$5.8\times 10^{-3}$  &$1.3\times 10^{-5}$    \\
    $ 3^3P_1\to 2^3D_1\to 2^3P_1\to \Upsilon(2S)$   &$9.0\times 10^{-3}$ &$18\%$  &11.5\%  & $1.9\times 10^{-4}$     \\
    $ 3^3P_1\to 2^3D_1\to 2^3P_0\to \Upsilon(1S)$   &$9.0\times 10^{-3}$ &$25\%$  &$2.2\times 10^{-3}$  &$5.0\times 10^{-6}$    \\
    $ 3^3P_1\to 2^3D_1\to 2^3P_1\to \Upsilon(1S)$   &$9.0\times 10^{-3}$ &$18\%$  &8.1\%  & $1.3\times 10^{-4}$     \\
    $ 3^3P_1\to 2^3D_1\to 1^3P_0\to \Upsilon(1S)$   &$9.0\times 10^{-3}$ &$15\%$  &1.76\%  &$2.4\times 10^{-5}$     \\
    $ 3^3P_1\to 2^3D_1\to 1^3P_1\to \Upsilon(1S)$   &$9.0\times 10^{-3}$ &$7\%$   &33.9\%  &$2.1\times 10^{-4}$      \\
    \hline
    $ 3^3P_1\to 2^3D_2\to 2^3P_1\to \Upsilon(2S)$   &$8.0\times 10^{-3}$ &$50\%$  &11.5\%  & $4.6\times 10^{-4}$     \\
    $ 3^3P_1\to 2^3D_2\to 2^3P_2\to \Upsilon(2S)$   &$8.0\times 10^{-3}$ &$16\%$  &11.0\%  & $1.4\times 10^{-4}$     \\
    $ 3^3P_1\to 2^3D_2\to 2^3P_1\to \Upsilon(1S)$   &$8.0\times 10^{-3}$ &$50\%$  &8.1\%  & $3.2\times 10^{-4}$     \\
    $ 3^3P_1\to 2^3D_2\to 2^3P_2\to \Upsilon(1S)$   &$8.0\times 10^{-3}$ &$16\%$  &9.5\%  &$1.2\times 10^{-4}$     \\
    $ 3^3P_1\to 2^3D_2\to 1^3P_1\to \Upsilon(1S)$   &$8.0\times 10^{-3}$ &$17\%$  &33.9\%  & $4.6\times 10^{-4}$     \\
    $ 3^3P_1\to 2^3D_2\to 1^3P_2\to \Upsilon(1S)$   &$8.0\times 10^{-3}$ &$5 \%$  &19.1\%  &$4.8\times 10^{-5}$     \\

\hline\hline
\end{tabular}
\end{center}
\end{table*}

\begin{table*}[htb]
\begin{center}
\caption{ Two-photon decay chains of $4^3S_1$. The combined branching fractions of the chain are defined by $\mathcal{B}=\mathcal{B}_1\times \mathcal{B}_2$ with $\mathcal{B}_1=\mathcal{B}[4^3S_1\to 3^3P_J \gamma]$, and $\mathcal{B}_2=\mathcal{B}[3^3P_J\to m^3S_{1}\gamma]$. }\label{4sp}
\begin{tabular}{c|c|c|ccccccccc}
\hline\hline
  Decay chain   &$\mathcal{B}_1 $$(10^{-4})$ & $\mathcal{B}_2$$(10^{-2})$ & $\mathcal{B}(10^{-6})$ \\
 \hline
    $4^3S_1\to 3^3P_2\to 1^3S_1  $   &$2.1 $ &$3.3$    & $6.9$              \\
    $4^3S_1\to 3^3P_1\to 1^3S_1  $   &$2.4$  &$5.4$    & $13$               \\
    $4^3S_1\to 3^3P_0\to 1^3S_1  $   &$1.7$  &$0.075$  & $0.13$               \\
    $4^3S_1\to 3^3P_2\to 2^3S_1  $   &$2.1 $ &$2.7$    &$5.7$               \\
    $4^3S_1\to 3^3P_1\to 2^3S_1  $   &$2.4$ &$4.8$     &$12$               \\
    $4^3S_1\to 3^3P_0\to 2^3S_1  $   &$1.7$ &$0.010$   &$0.017$               \\
    $4^3S_1\to 3^3P_2\to 3^3S_1  $   &$2.1$ &$4.4$     &$9.2$               \\
    $4^3S_1\to 3^3P_1\to 3^3S_1  $   &$2.4$ &$8.8$     &$21$               \\
    $4^3S_1\to 3^3P_0\to 3^3S_1  $   &$1.7$ &$0.032$   &$0.054$               \\
\hline\hline
\end{tabular}
\end{center}
\end{table*}

\section*{  Acknowledgements }

This work is supported, in part, by the National Natural Science
Foundation of China (Grants No. 11075051 and No. 11375061), and the
Hunan Provincial Natural Science Foundation (Grant No. 13JJ1018).


\section*{Appendix}

The method for solving Eq.(\ref{sdg}) is outlined as follows.
Equation (\ref{sdg}) can be rewritten as
\begin{eqnarray}
\frac{d^2u(r)}{dr^2}=T(r)u(r),
\end{eqnarray}
with $T(r)=-2\mu_R \left[E-V_{b\bar{b}}(r)-\frac{L(L+1)}{2\mu_R r^2}\right]$.
According to the Gowell central difference method,
we have~\cite{Haicai}
\begin{eqnarray}\label{ads}
u(r_{i+1})=\frac{[2+\frac{5}{6}h^2T(r_i)]u(r_{i})-[1-\frac{1}{12}h^2T(r_{i-1})]u(r_{i-1})}{1-\frac{1}{12}h^2T(r_{i+1})},
\end{eqnarray}
with $r_i=ih \ (i=0,1,2\cdot\cdot\cdot)$. The starting conditions
of the above equation are
\begin{eqnarray}\label{ads2}
u(0)=0,\ \ \ u(h)=h^{L+1},\nonumber\\
T(0)u(0)=\lim_{r\to 0}\frac{L(L+1)}{r^2}r^{L+1}=2\delta_{L1}.
\end{eqnarray}
Thus, for a given binding energy $E$, in terms of Eq.(\ref{ads}),
the radial wave function $u(r)$ can be calculated
from the center ($r=0$) towards the outside ($r\to \infty$) point by point.

Finally, to determine the binding energy $E$, we adopt the following method.
As we know if $E_0$ is a trial value near the eigenvalue of the binding energy $E$,
the asymptotic form of the numerical solution of the radial wave function $u(r,E_0)$ at large $r$
is given by the linear combination of the regular solution $g(E_0)e^{-k_0r}$
and irregular solution $f(E_0)e^{+k_0r}$ with $k_0^2=2\mu E_0$. Thus, we can take
the radial wave function $u(r,E_0)$ at large $r$ as~\cite{Haicai}
\begin{equation}
u(r,E_0)=f(E_0)e^{+k_0r},
\end{equation}
Similarly, for another trial value $E_1$, we have
\begin{equation}
u(r,E_1)=f(E_1)e^{+k_1r},
\end{equation}
with $k_1^2=2\mu E_1$. If $f(E)$ is an analytic function,
we can expand $f(E_1)$ as
\begin{equation}\label{e82}
f(E_1)=f(E_0)+f'(E_0)(E_1-E_0)+\cdot\cdot\cdot.
\end{equation}
If $|E_1-E_0|$ is small enough, we can only keep the first two terms.
Then, we have
\begin{equation}
f'(E_0)=\frac{f(E_1)-f(E_0)}{E_1-E_0}=\frac{u(r,E_1)e^{-k_1 r}-u(r,E_0)e^{-k_0 r}}{E_1-E_0}.
\end{equation}
Note that, if $E_1$ is just the eigenvalue of the binding energy $E$,
$f(E_1)$ should be zero. Thus, from Eq.(\ref{e82}) we have
\begin{equation}
E=E_0-f(E_0)/f'(E_0)
\end{equation}
In the numerical calculations, the recurrence method is used to calculate
the eigenvalue $E$. Letting $E_1\to E_0$, $u(r,E_1)\to u(r,E_0)$ and $E\to E_1$,
then we calculate new $u(r,E_1)$ and new $E$ with Eqs.(\ref{ads}) and (\ref{ads2}).
The recurrence is stopped when $|E-E_0|\leq \epsilon$, where $\epsilon$ stands for the accuracy that we need.



\begin{thebibliography}{99}


\bibitem{Godfrey:1985xj}
  S.~Godfrey and N.~Isgur,
  Mesons in a Relativized Quark Model with Chromodynamics,
  Phys.\ Rev.\ D {\bf 32}, 189 (1985).

\bibitem{Eichten:1978tg}
  E.~Eichten, K.~Gottfried, T.~Kinoshita, K.~D.~Lane and T.~M.~Yan,
  Charmonium: The Model,
  Phys.\ Rev.\ D {\bf 17}, 3090 (1978)
  Erratum: [Phys.\ Rev.\ D {\bf 21}, 313 (1980)].

\bibitem{Garmash:2015xea}
  A.~Garmash,
  Bottomonium Studies at Belle,
  EPJ Web Conf.\  {\bf 96}, 01014 (2015).

\bibitem{Eichten:2007qx}
  E.~Eichten, S.~Godfrey, H.~Mahlke and J.~L.~Rosner,
  Quarkonia and their transitions,
  Rev.\ Mod.\ Phys.\  {\bf 80}, 1161 (2008).



\bibitem{Brambilla:2004wf}
  N.~Brambilla {\it et al.} [Quarkonium Working Group Collaboration],
  Heavy quarkonium physics,
  hep-ph/0412158.


\bibitem{Brambilla:2010cs}
  N.~Brambilla {\it et al.},
  Heavy quarkonium: progress, puzzles, and opportunities,
  Eur.\ Phys.\ J.\ C {\bf 71}, 1534 (2011).


\bibitem{Bevan:2014iga}
  A.~J.~Bevan {\it et al.} [BaBar and Belle Collaborations],
  The Physics of the $B$ Factories,
  Eur.\ Phys.\ J.\ C {\bf 74}, 3026 (2014).


\bibitem{PDG}
  K.~A.~Olive {\it et al.} [Particle Data Group Collaboration],
  Review of Particle Physics,
  Chin.\ Phys.\ C {\bf 38}, 090001 (2014).



\bibitem{Gupta:1984jb}
  S.~N.~Gupta, S.~F.~Radford and W.~W.~Repko,
  $b\bar{b}$ Spectroscopy,
  Phys.\ Rev.\ D {\bf 30}, 2424 (1984).

\bibitem{Kwong:1988ae}
  W.~Kwong and J.~L.~Rosner,
  $D$ Wave Quarkonium Levels of the $\Upsilon$ Family,
  Phys.\ Rev.\ D {\bf 38}, 279 (1988).

\bibitem{Chao:2009}
  B.~Q.~Li and K.~T.~Chao,
  Bottomonium Spectrum with Screened Potential,
  Commun.\ Theor.\ Phys.\  {\bf 52}, 653 (2009).

\bibitem{Godfrey:2015dia}
  S.~Godfrey and K.~Moats,
  Bottomonium Mesons and Strategies for their Observation,
  Phys.\ Rev.\ D {\bf 92}, 054034 (2015).


\bibitem{Segovia:2016xqb}
  J.~Segovia, P.~G.~Ortega, D.~R.~Entem and F.~Fern\'{a}ndez,
  Bottomonium spectrum revisited,
  Phys.\ Rev.\ D {\bf 93}, 074027 (2016).

\bibitem{Barducci:2016wze}
  A.~Barducci, R.~Giachetti and E.~Sorace,
  Relativistic two-body calculation of $b\bar{b}$-mesons radiative decays,
  arXiv:1604.08043 [hep-ph].

\bibitem{Wei-Zhao:2013sta}
  Wei-Zhao Tian, Lu Cao, You-Chang Yang and Hong Chen,
  Bottomonium states versus recent experimental observations in the QCD-inspired potential model,
  Chin.\ Phys.\ C {\bf 37}, 083101 (2013).

\bibitem{Ebert:2003}
  D.~Ebert, R.~N.~Faustov and V.~O.~Galkin,
  Properties of heavy quarkonia and $B_c$ mesons in the relativistic quark model,
  Phys.\ Rev.\ D {\bf 67}, 014027 (2003).

\bibitem{Akbar:2015evy}
  N.~Akbar, M.~A.~Sultan, B.~Masud and F.~A.~Sultan,
  Higher Hybrid Bottomonia in an Extended Potential Model,
  arXiv:1511.03632 [hep-ph].

\bibitem{Hughes:2015dba}
  C.~Hughes, R.~J.~Dowdall, C.~T.~H.~Davies, R.~R.~Horgan, G.~von Hippel and M.~Wingate,
  Hindered M1 Radiative Decay of $\Upsilon(2S)$ from Lattice NRQCD,
  Phys.\ Rev.\ D {\bf 92}, 094501 (2015).


\bibitem{Lewis:2011ti}
  R.~Lewis and R.~M.~Woloshyn,
  Excited Upsilon Radiative Decays,
  Phys.\ Rev.\ D {\bf 84}, 094501 (2011).

\bibitem{Becirevic:2014rda}
  D.~Becirevi, M.~Kruse and F.~Sanfilippo,
  Lattice QCD estimate of the $\eta_{c}(2S)\to J/\psi \gamma$  decay rate,
  JHEP {\bf 1505}, 014 (2015).

\bibitem{Baker:2015xma}
  M.~Baker, A.~A.~Penin, D.~Seidel and N.~Zerf,
  Bottomonium Hyperfine Splitting on the Lattice and in the Continuum,
  Phys.\ Rev.\ D {\bf 92}, 054502 (2015).

\bibitem{DeFazio:2008xq}
  F.~De Fazio,
  Radiative transitions of heavy quarkonium states,
  Phys.\ Rev.\ D {\bf 79}, 054015 (2009)
  Erratum: [Phys.\ Rev.\ D {\bf 83}, 099901 (2011)].

\bibitem{Brambilla:2005zw}
  N.~Brambilla, Y.~Jia and A.~Vairo,
  Model-independent study of magnetic dipole transitions in quarkonium,
  Phys.\ Rev.\ D {\bf 73}, 054005 (2006).

\bibitem{Brambilla:2012be}
  N.~Brambilla, P.~Pietrulewicz and A.~Vairo,
  Model-independent Study of Electric Dipole Transitions in Quarkonium,
  Phys.\ Rev.\ D {\bf 85}, 094005 (2012).

\bibitem{Pineda:2013lta}
  A.~Pineda and J.~Segovia,
  Improved determination of heavy quarkonium magnetic dipole transitions in potential nonrelativistic QCD,
  Phys.\ Rev.\ D {\bf 87}, 074024 (2013).

\bibitem{Ferretti:2013vua}
  J.~Ferretti and E.~Santopinto,
  Higher mass bottomonia,
  Phys.\ Rev.\ D {\bf 90}, 094022 (2014).

\bibitem{Ferretti:2014xqa}
  J.~Ferretti, G.~Galat¨¤ and E.~Santopinto,
  Quark structure of the $X(3872)$ and $\chi_b(3P)$ resonances,
  Phys.\ Rev.\ D {\bf 90}, 054010 (2014).

\bibitem{Lu:2016mbb}
  Y.~Lu, M.~N.~Anwar and B.~S.~Zou,
  Coupled-Channel Effects for the Bottomonium with Realistic Wave Functions,
  Phys.\ Rev.\ D {\bf 94}, 034021 (2016).


\bibitem{Choi:2007se}
  H.~M.~Choi,
  Decay constants and radiative decays of heavy mesons in light-front quark model,
  Phys.\ Rev.\ D {\bf 75}, 073016 (2007).

\bibitem{Ke:2010vn}
  H.~W.~Ke, X.~Q.~Li, Z.~T.~Wei and X.~Liu,
  Re-Study on the wave functions of $\Upsilon(nS)$ states in LFQM and the radiative decays of $\Upsilon(nS)\to \eta_b\gamma$,
  Phys.\ Rev.\ D {\bf 82}, 034023 (2010).

\bibitem{Ke:2010tk}
  H.~W.~Ke, X.~Q.~Li and X.~Liu,
  Study on the Radiative Decays of $\Upsilon(nS)\to \eta_b\gamma$,
  arXiv:1002.1187 [hep-ph].

\bibitem{Ke:2013zs}
  H.~W.~Ke, X.~Q.~Li and Y.~L.~Shi,
  The radiative decays of $0^{++}$ and $1^{+-}$ heavy mesons,
  Phys.\ Rev.\ D {\bf 87}, 054022 (2013).

\bibitem{Li:2009zu}
  B.~Q.~Li and K.~T.~Chao,
  Higher Charmonia and X,Y,Z states with Screened Potential,
  Phys.\ Rev.\ D {\bf 79}, 094004 (2009).

\bibitem{ChaoKT90}
K.T. Chao and J.H. Liu, in Proceedings of the Workshop
on Weak Interactions and CP Violation, Beijing, August
22-26, 1989, edited by T. Huang and D.D. Wu, World
Scientific (Singapore, 1990) p.109-p.117.

\bibitem{ChaoKT93}
Y. B. Ding, K. T. Chao and D. H. Qin,
Screened $Q\bar{Q}$ potential and spectrum of heavy quarkonium,
Chin. Phys. Lett.{\bf 10}, 460 (1993).

\bibitem{Laermann:1986pu}
  E.~Laermann, F.~Langhammer, I.~Schmitt and P.~M.~Zerwas,
  The Interquark Potential: SU(2) Color Gauge Theory With Fermions,
  Phys.\ Lett.\ B {\bf 173}, 437 (1986).

\bibitem{Born:1989iv}
  K.~D.~Born, E.~Laermann, N.~Pirch, T.~F.~Walsh and P.~M.~Zerwas,
 Hadron Properties in Lattice {QCD} With Dynamical Fermions,
  Phys.\ Rev.\ D {\bf 40}, 1653 (1989).

\bibitem{Deng:2015bva}
  W.~J.~Deng, L.~Y.~Xiao, L.~C.~Gui and X.~H.~Zhong,
  Radiative transitions of charmonium states,
  arXiv:1510.08269 [hep-ph].

\bibitem{Li:1994cy}
  Zhenping~Li,
  Threshold pion photoproduction of nucleons in the chiral quark model,
  Phys.\ Rev.\ D {\bf 50}, 5639 (1994).


\bibitem{Li:1995si}
  Zhenping Li,
  The Kaon photoproduction of nucleons in the chiral quark model,
  Phys.\ Rev.\ C {\bf 52}, 1648 (1995).


\bibitem{Li:1997gda}
  Zhenping Li, H.~X.~Ye and M.~H.~Lu,
  An unified approach to pseudoscalar meson photoproductions off  nucleons in
  the quark model,
  Phys.\ Rev.\  C {\bf 56}, 1099 (1997).

\bibitem{Zhao:2001kk}
  Q.~Zhao,
  Eta-prime photoproduction near threshold,
  Phys.\ Rev.\ C {\bf 63}, 035205 (2001).



\bibitem{Saghai:2001yd}
  B.~Saghai and Zhenping Li,
  Quark model study of the eta photoproduction: Evidence for a new $S_{11}$ resonance?,
  Eur.\ Phys.\ J.\ A {\bf 11}, 217 (2001).

\bibitem{Zhao:2002id}
  Q.~Zhao, J.~S.~Al-Khalili, Z.~P.~Li and R.~L.~Workman,
  Pion photoproduction on the nucleon in the quark model,
  Phys.\ Rev.\ C {\bf 65}, 065204 (2002).

\bibitem{He:2008ty}
  J.~He, B.~Saghai and Z.~Li,
  Study of $\eta$ photoproduction on the proton in a chiral constituent quark approach via one-gluon-exchange model,
  Phys.\ Rev.\ C {\bf 78}, 035204 (2008).

\bibitem{He:2008uf}
  J.~He and B.~Saghai,
  Combined study of $\gamma p \to \eta p$ and $\pi^- p \to \eta n$ in a chiral constituent quark approach,
  Phys.\ Rev.\ C {\bf 80}, 015207 (2009).

\bibitem{He:2010ii}
  J.~He and B.~Saghai,
  $\eta$ production off the proton in a Regge-plus-chiral quark approach,
  Phys.\ Rev.\ C {\bf 82}, 035206 (2010).

\bibitem{Zhong:2011ti}
  X.~H.~Zhong and Q.~Zhao,
  $\eta$ photoproduction on the quasi-free nucleons in the chiral quark model,
  Phys.\ Rev.\ C {\bf 84}, 045207 (2011).

\bibitem{Zhong:2011ht}
  X.~H.~Zhong and Q.~Zhao,
  $\eta'$ photoproduction on the nucleons in the quark model,
  Phys.\ Rev.\ C {\bf 84}, 065204 (2011).

\bibitem{Xiao:2015gra}
  L.~Y.~Xiao, X.~Cao and X.~H.~Zhong,
  Neutral pion photoproduction on the nucleon in a chiral quark model,
  Phys.\ Rev.\ C {\bf 92}, 035202 (2015).

\bibitem{Barnes:2005pb}
  T.~Barnes, S.~Godfrey and E.~S.~Swanson,
  Higher charmonia,
  Phys.\ Rev.\ D {\bf 72}, 054026 (2005).

\bibitem{DF1982}
  D.~Flamm and F.~Sch\"{o}ber,
  Itroduction to the quark model of elementary particle,
  Volume 1: Quantum numbers, gauge theory and hadron spectroscopy,
  Gordon and Breach science publishers (1982).



\bibitem{Haicai}
 Chong-Hai Cai and Lei Li,
  Radial equation of bound state and binding energies of $\Xi^-$ hypernuclei,
  High Energy Physics and Nuclear Physics {\bf 27}, 1005 (2003).

\bibitem{Abdesselam:2016xbr}
  A.~Abdesselam {\it et al.},
  Study of $\chi_{bJ}(1P)$ Properties in the Radiative $\Upsilon(2S)$ Decays,
  arXiv:1606.01276 [hep-ex].

\bibitem{delAmoSanchez:2010kz}
  P.~del Amo Sanchez {\it et al.} [BaBar Collaboration],
  Observation of the $\Upsilon(1^3D_J)$ Bottomonium State through Decays to $\pi^+\pi^-\Upsilon(1S)$,
  Phys.\ Rev.\ D {\bf 82}, 111102 (2010).

\bibitem{Cawlfield:2005ra}
  C.~Cawlfield {\it et al.} [CLEO Collaboration],
 Experimental study of $\chi_b(2P) \to \pi \pi \chi_b(1P)$,
  Phys.\ Rev.\ D {\bf 73}, 012003 (2006).

\bibitem{Mizuk:2012pb}
  R.~Mizuk {\it et al.} [Belle Collaboration],
  Evidence for the $\eta_b(2S)$ and observation of $h_b(1P) \to \eta_b(1S) \gamma$ and $h_b(2P) \to \eta_b(1S) \gamma$,
  Phys.\ Rev.\ Lett.\  {\bf 109}, 232002 (2012).

\bibitem{Badalian:2012mb}
  A.~M.~Badalian and B.~L.~G.~Bakker,
  Dominant spin-orbit effects in radiative decays $\Upsilon(3S)\to \gamma\chi_{bJ}(1P))$,
  Phys.\ Rev.\ D {\bf 86}, 074001 (2012).

\bibitem{Aad:2011ih}
  G.~Aad {\it et al.} [ATLAS Collaboration],
  Observation of a new $\chi_b$ state in radiative transitions to $\Upsilon(1S)$ and $\Upsilon(2S)$ at ATLAS,
  Phys.\ Rev.\ Lett.\  {\bf 108}, 152001 (2012).


\bibitem{Abazov:2012gh}
  V.~M.~Abazov {\it et al.} [D0 Collaboration],
  Observation of a narrow mass state decaying into $\Upsilon(1S) \gamma$ in $p\bar{p}$ collisions at $\sqrt{s} = 1.96$ TeV,
  Phys.\ Rev.\ D {\bf 86}, 031103 (2012).

\bibitem{Aaij:2014hla}
  R.~Aaij {\it et al.} [LHCb Collaboration],
  Measurement of the $\chi_b(3P)$ mass and of the relative rate of $\chi_{b1}(1P)$ and $\chi_{b2}(1P)$ production,
  JHEP {\bf 1410}, 88 (2014).


\bibitem{Aaij:2014caa}
  R.~Aaij {\it et al.} [LHCb Collaboration],
  Study of $\chi _{b}$ meson production in $pp$ collisions at $\sqrt{s}=7$ and $8{\mathrm {\,TeV}} $ and observation of the decay $\chi _{b}(3P)\to \Upsilon (3S)\gamma$,
  Eur.\ Phys.\ J.\ C {\bf 74}, 3092 (2014).






\end{thebibliography}
\end{document}